\documentclass[3p]{elsarticle}

\AtBeginDocument{}
\usepackage{cmap}
\usepackage{bm}
\usepackage[T1]{fontenc}
\usepackage{float}
\usepackage{tikz}
\usepackage{amssymb,amsbsy,amsmath,amsfonts,amssymb,amscd}
\usepackage{verbatim}
\usepackage{wrapfig}
\usepackage{algorithmic}
\usepackage[]{algorithm2e}
\usepackage{graphicx}
\usepackage{graphics}
\usepackage{stmaryrd}
\usepackage[small]{caption}
\usepackage{subcaption}
\usepackage{hyperref}
\usepackage{color}
\usepackage{amsthm}  
\usepackage{tabularx}
\usepackage[latin1]{inputenc}
\usepackage{dcolumn}
\restylefloat{table}
\usepackage{booktabs}
\usepackage{xcolor,colortbl}
\usepackage{amssymb}
\usepackage{amsfonts}
\usepackage{graphicx}
\usepackage{parskip}

\newcommand{\norm}[1]{\left|#1\right|}

\graphicspath{{figures_lowSize/}}

\begin{document}

\begin{frontmatter}

\title{A physics-based shock capturing method for large-eddy simulation}

\author[add1,add2]{P. Fernandez\corref{1}}
\ead{pablof@mit.edu}
\author[add1,add2]{N.~C. Nguyen}
\ead{cuongng@mit.edu}
\author[add1,add2]{J. Peraire}
\ead{peraire@mit.edu}
\address[add1]{Department of Aeronautics and Astronautics, Massachusetts Institute of Technology, USA.}
\address[add2]{Center for Computational Engineering, Massachusetts Institute of Technology, USA.}
\cortext[1]{Corresponding author}

\begin{abstract}

We present a shock capturing method for large-eddy simulation of turbulent flows. 
The proposed method relies on physical mechanisms to resolve and smooth sharp unresolved flow features that may otherwise lead to numerical instability, such as shock waves and under-resolved thermal and shear layers. 
To that end, we devise various sensors to detect when and where the shear viscosity, bulk viscosity and thermal conductivity of the fluid do not suffice to stabilize the numerical solution. In such cases, the fluid viscosities are selectively increased to ensure the cell P\'eclet number is of order $1$ so that these flow features can be well represented with the grid resolution. 
Although the shock capturing method is devised in the context of discontinuous Galerkin methods, it can be used with other discretization schemes. The performance of the method is illustrated through numerical simulation of external and internal flows in transonic, supersonic, and hypersonic regimes. For the problems considered, the shock capturing method performs robustly, provides sharp shock profiles, and has a small impact on the resolved turbulent structures. 
These three features are critical to enable robust and accurate large-eddy simulations of shock flows.

\end{abstract}


\end{frontmatter}

%

\section{\label{s:introduction}Introduction}

The use of computational fluid dynamics (CFD) in industry is severely limited by the inability to accurately and reliably simulate complex turbulent flows. This is partly due to the current numerical technologies adopted by industry practitioners, that still rely on steady-tailored techniques, in conjunction with low-order numerical methods. In fact, the majority of CFD codes have first or second order spatial accuracy and are based on Reynolds-Averaged Navier-Stokes (RANS) equations or, more recently, detached-eddy simulation (DES). The use of high-fidelity computer-aided design is still very limited, with large-eddy simulation (LES) largely confined in the research and development branches of industry, or in academia. However, with the increase in computing power, LES is becoming a feasible approach to model complex industrial turbulent flows. At the same time, discontinuous Galerkin (DG) methods are gaining attention for large-eddy simulation of these flows, as they allow for high-order discretizations on complex geometries and unstructured meshes. This is critical to accurately propagate small-scale, small-magnitude features, such as in transitional and turbulent flows, over the complex three-dimensional geometries commonly encountered in industrial applications. In addition, DG methods are well suited to emerging computing architectures, including graphics processing units (GPUs) and other many-core architectures, due to their high flop-to-communication ratio. The use of DG methods for LES is being further encouraged by successful numerical predictions \cite{Beck:14,Fernandez:16a,Fernandez:17a,Frere:15,Gassner:13,Murman:16,Renac:15,Uranga:11,Wiart:15}.

Large-eddy simulations are, by definition, under-resolved computations. As customary, we use the term `under-resolved' to refer to simulations in which the exact solution contains scales that are smaller than the Nyquist wavenumber of the grid (the so-called subgrid scales, briefly SGS) and thus cannot be captured with the grid resolution. 
Two types of under-resolved features can exist in LES. ({\em i}) Small-magnitude features that are lost in the numerical solution due to the filter introduced by the numerics, such as the turbulent structures that are smaller than the grid size. This type of subgrid scales are accounted for, and stabilized, by the implicit or explicit SGS model. As customary in the literature, they will be referred to simply as `subgrid scales'. 
({\em ii}) Large-magnitude, sharp features that remain in the discrete solution. 
A number of such sharp features may appear in LES, especially in transonic, supersonic and hypersonic flows, including shock waves, contact discontinuities, strong thermal gradients, and thin shear layers. We shall refer to them as `sharp subgrid-scale features' or simply `sharp features'. Insufficient resolution to capture sharp features usually leads to Gibbs oscillations and inaccurate results, and may even lead to nonlinear instability and simulation breakdown.

Despite the large number of works on shock capturing since the dawn of computational fluid dynamics, numerical simulation of turbulent shock flows remains a challenging problem \cite{Johnsen:2010,Mani:09,Slotnick:14:NASA}, particularly for high-order discretization schemes \cite{Hillewaert:2016}. First, many shock capturing methods have been developed for steady-state or inviscid problems, but their extension to unsteady viscous flows is not straightforward. Second, the majority of the existing methods are by construction not able to stabilize sharp features other than shock waves. This compromises robustness and limits the applicability of the method to some particular types of flows. In order to enable large-eddy simulation of transonic, supersonic and hypersonic flows, a method to detect and stabilize all sources of numerical instability is required. While this is referred to as a {\it shock capturing} method for consistency with the common terminology in the literature, we emphasize the need to deal with other sharp subgrid-scale features in LES.

Prior to describing the proposed shock capturing method, we present an overview of strategies in the literature for the detection and stabilization of shock waves and other sharp features. As for shock detection, perhaps the most popular approach is to take advantage of the strong compression that a fluid undergoes across a shock wave and use the divergence of the velocity field as a shock sensor \cite{Barter:2010,Moro:2016,Nguyen:11}. An assessment of dilatation-based shock detection methods is presented in \cite{Yu:2017}. In our experience, the existing methods in this category provide non-oscillatory shocks for steady flows, such as laminar and Reynolds-averaged turbulent flows, but fail for unsteady turbulent flows. Alternatively, a number of methods rely on the non-smoothness of the numerical solution to detect shocks as well as other sharp features \cite{Cook:2005,Kawai:2008,Kawai:2010,Klockner:2011,Krivodonova:2004,Olson:2013,Persson:2006,Persson:2013}. Among them, the sensor by Krivodonova {\it et al.} \cite{Krivodonova:2004} takes advantage of the theoretical convergence rate of DG schemes for smooth solutions in order to detect discontinuities. By construction, this sensor is limited to high-order DG methods, hyperbolic systems of conservation laws such as the Euler equations, and stabilization mechanisms that do not introduce artificial viscosity. The shock sensor by Persson {\it et al.} \cite{Persson:2006,Persson:2013} is based on the decay rate of the coefficients of the DG polynomial approximation. Like the sensor by Krivodonova, it requires accuracy orders beyond about $5$ to provide accurate results. Other approaches that rely on high-order derivatives of the solution include \cite{Cook:2005,Kawai:2008,Kawai:2010,Olson:2013}, but again apply only to schemes for which such derivatives can be accurately computed, such as spectral-type methods and high-order finite difference methods on structured meshes and simple geometries. Also, most methods based on the smoothness of the numerical solution involve user-defined parameters that are flow-dependent and usually hard to tune for new problems. This compromises the adoption of these methods in industry due to the high robustness and flexibility that these applications require.

Regarding stabilization of sharp features, most methods lie within one of the following two categories: Limiters and artificial viscosity. Limiters, in the form of flux limiters \cite{Burbeau:2001,Cockburn:1989,Krivodonova:2007} or solution limiters \cite{Cockburn:1998,Krivodonova:2004,Lv:2015,Luo:2007,Qiu:2005,Sonntag:2017,Zhu:2008}, are in general not well suited for implicit time integration schemes and additionally pose challenges for high-order methods on complex geometries. As for artificial viscosity methods, Laplacian-based \cite{Hartmann:2013,Lv:2016,Nguyen:11,Moro:2016,Persson:2006,Persson:2013} and physics-based \cite{Abbassi:2014,Cook:2005,Kawai:2008,Kawai:2010,Olson:2013,Persson:2006,Shebalin:1993} approaches have been proposed. An assessment of artificial viscosity methods for LES is presented in \cite{Mani:09}. In general, these methods perform poorly for unsteady flows and/or require accurate high-order derivatives of the numerical solution. 
In this paper, we present a shock capturing method for large-eddy simulation that aims to address the limitations above. Our approach comprises physics-based sensors to detect shock waves and other sharp features, as well as physics-based artificial viscosities to stabilize them. Although our approach can be implemented with other numerical schemes, the hybridized discontinuous Galerkin methods are considered for illustration purposes. 
The performance of the method is examined through numerical simulation of external and internal flows in transonic, supersonic, and hypersonic regimes. A comparative study is conducted between the proposed approach and a Laplacian-based approach, widely used in the DG community, in order to illustrate the importance of using physical viscosities, as opposed to Laplacian viscosities, for large-eddy simulation of turbulent flows.

The remainder of the paper is organized as follows. In Section \ref{s:discretization}, we present the numerical discretization of the Navier-Stokes equations. Sections \ref{s:sensors} and \ref{s:AV} describe the sensors to detect sharp features and the procedure to stabilize the numerical scheme, respectively. The performance of the shock capturing method for a number of flow conditions is illustrated in Section \ref{s:results}. We conclude the paper with some remarks and future work in Section \ref{s:conclusions}.

\section{\label{s:discretization}Flow discretization}

\subsection{Governing equations}

Let $t_f > 0$ be a final time and let $\Omega \subset \mathbb{R}^d, \, 1 \leq d \leq 3$ be an open, connected and bounded physical domain with Lipschitz 
boundary $\partial \Omega$. We consider the unsteady, compressible Navier-Stokes equations written in conservation form as
\begin{subequations}
\label{e:NS}
\begin{alignat}{2}
\label{e:ns0}
\displaystyle \bm{q} - \nabla \bm{u} = 0 & \mbox{ ,} \qquad \mbox{in } \Omega \times (0, t_f) \mbox{ ,}   \\
\label{e:ns1}
\displaystyle \frac{\partial  \bm{u}}{\partial t} +  \nabla  \cdot  \bm{F}(\bm{u}) +  \nabla  \cdot  \bm{G}(\bm{u} , \bm{q}) = 0 & \mbox{ ,} \qquad \mbox{in } \Omega \times (0, t_f) \mbox{ ,}   \\
\label{e:ns2}
\bm{B}(\bm{u} , \bm{q}) = 0 & \mbox{ ,} \qquad \mbox{on } \partial \Omega \times (0,t_f) \mbox{ ,} \\
\label{e:ns3}
\bm{u} - \bm{u}_0 = 0 & \mbox{ ,} \qquad \mbox{on } \Omega \times \{ 0 \} \mbox{ .} 
\end{alignat}
\end{subequations}
Here, $\bm{u} = (\rho, \rho v_{j}, \rho E), \ j=1,...,d$ is the $m$-dimensional ($m = d + 2$) vector of conserved quantities, $\bm{u}_0$ is an initial condition, $\bm{B}(\bm{u},\bm{q})$ is a boundary operator, and $\bm{F}(\bm{u})$ and $\bm{G}(\bm{u} , \bm{q})$ are the inviscid and viscous fluxes of dimensions $m \times d$,
\begin{equation}
\label{e:flux}
\bm{F}(\bm{u}) = \left( \begin{array}{c}
\rho v_j \\
\rho v_i v_j + \delta_{ij} p \\
 v_j (\rho E + p)
\end{array}
\right) \mbox{ ,} \qquad
\bm{G}(\bm{u},\bm{q}) = - \left( \begin{array}{c}
0 \\
\tau_{ij}  \\
v_i \tau_{ij} - f_j
\end{array}
\right) \mbox{ ,} \qquad i , j = 1 , \dots , d \mbox{ ,} 
\end{equation}
where $p$ denotes the thermodynamic pressure, $\tau_{ij}$ the viscous stress tensor, $f_j$ the heat flux, and $\delta_{ij}$ is the Kronecker delta. For a calorically perfect gas in thermodynamic equilibrium, $p = (\gamma - 1) \, \big( \rho E - \rho \, \norm{\bm{v}}^2 / 2 \big)$, where $\gamma = c_p / c_v > 1$ is the ratio of specific heats and in particular $\gamma \approx 1.4$ for air. $c_p$ and $c_v$ are the specific heats at constant pressure and volume, respectively. For a Newtonian fluid with the Fourier's law of heat conduction, the viscous stress tensor and heat flux are given by
\begin{equation}
\label{closures}
\tau_{ij} = \mu \, \bigg( \frac{\partial v_i}{\partial x_j} + \frac{\partial v_j}{\partial x_i} - \frac{2}{3}\frac{\partial v_k}{\partial x_k}\delta_{ij} \bigg) + \beta \, \frac{\partial v_k}{\partial x_k}\delta_{ij} ,  \qquad \qquad \qquad f_j = - \, \kappa \, \frac{\partial T}{\partial x_j} , 
\end{equation}
where $T$ denotes temperature, $\mu$ the dynamic (shear) viscosity, $\beta$ the bulk viscosity, $\kappa = c_p \, \mu / Pr$ the thermal conductivity, and $Pr$ the Prandtl number. In particular, $Pr \approx 0.71$ for air, and additionally $\beta = 0$ under the Stokes' hypothesis. 

The numerical examples in Section \ref{s:results} include inviscid flows, governed by the unsteady compressible Euler equations. The Euler equations are obtained by dropping the viscous flux in Eq. \eqref{e:ns1}.

\subsection{Numerical discretization}

We consider the hybridized discontinuous Galerkin (DG) methods \cite{Fernandez:17a}, which generalize the Hybridizable DG (HDG) \cite{Nguyen:12,Peraire:10}, the Embedded DG (EDG) \cite{Peraire:11} and the Interior Embedded DG (IEDG) \cite{Fernandez:16a} methods, for the spatial discretization of the unsteady compressible Navier-Stokes equations. The hybridized DG discretization reads as follows: Find $\big( \bm{q}_h(t),\bm{u}_h(t), \widehat{\bm{u}}_h(t) \big) \in \bm{\mathcal{Q}}_h^k \times \bm{\mathcal{V}}_h^k \times \bm{\mathcal{M}}_h^k$ such that
\begin{subequations}
\label{IEDG}
\begin{alignat}{2}
\label{IEDGa}
\big( \bm{q}_h, \bm{r} \big) _{\mathcal{T}_h} + \big( \bm{u}_h, \nabla \cdot \bm{r} \big)  _{\mathcal{T}_h} -  \big< \widehat{\bm{u}}_h, \bm{r} \cdot \bm{n} \big> _{\partial \mathcal{T}_h}  & =  0 \mbox{ ,} \\
\label{IEDGb}
\Big( \frac{\partial \, \bm{u}_h}{\partial t}, \bm{w} \Big)_{\mathcal{T}_h} - \Big( \bm{F}(\bm{u}_h) + \bm{G}(\bm{u}_h,\bm{q}_h) , \nabla \bm{w} \Big) _{\mathcal{T}_h}  +  \left\langle \widehat{\bm{f}}_h(\widehat{\bm{u}}_h,\bm{u}_h) + \widehat{\bm{g}}_h(\widehat{\bm{u}}_h,\bm{u}_h,\bm{q}_h), \bm{w} \right\rangle_{\partial \mathcal{T}_h}  & = 0 \mbox{ ,} \\
\label{IEDGc}
\left\langle \widehat{\bm{f}}_h(\widehat{\bm{u}}_h,\bm{u}_h) + \widehat{\bm{g}}_h(\widehat{\bm{u}}_h,\bm{u}_h,\bm{q}_h), \bm{\mu} \right\rangle_{\partial \mathcal{T}_h \backslash \partial \Omega} + \left\langle \widehat{\bm{b}}_h(\widehat{\bm{u}}_h,\bm{u}_h,\bm{q}_h), \bm{\mu} \right\rangle_{\partial \Omega} & =  0 \mbox{ ,} \\
\intertext{for all $(\bm{r},\bm{w}, {\bm{\mu}}) \in \bm{\mathcal{Q}}^k_h \times \bm{\mathcal{V}}^k_h \times \bm{\mathcal{M}}_{h}^k$ and all $t \in (0,t_f)$, as well as}
\label{IEDGd}
\big( \bm{u}_{h}|_{t=0} - \bm{u}_0 , \bm{w} \big) _{\mathcal{T}_h} & =  0 \mbox{ ,} 
\end{alignat}
\end{subequations}
for all $\bm{w} \in \bm{\mathcal{V}}^k_h$. The finite element spaces and inner products above are described in \ref{s:DGnotation}. 
The inviscid and viscous numerical fluxes, $\widehat{\bm{f}}_h$ and $\widehat{\bm{g}}_h$, are defined as
\begin{subequations}
\label{numericalFlux}
\begin{alignat}{2}
\label{e:numericalInviscidFlux}
& \widehat{\bm{f}}_h(\widehat{\bm{u}}_h , \bm{u}_h ) = \bm{F}(\widehat{\bm{u}}_h) \cdot \bm{n} + \bm{\sigma}(\widehat{\bm{u}}_h , \bm{u}_h ; \bm{n}) \cdot ( \bm{u}_h - \widehat{\bm{u}}_h ) \mbox{ ,} \\
\label{e:numericalViscousFluxNS}
& \widehat{\bm{g}}_h(\widehat{\bm{u}}_h , \bm{u}_h , \bm{q}_h ) = \bm{G}(\widehat{\bm{u}}_h , \bm{q}_h) \cdot \bm{n} \mbox{ ,} 
\end{alignat}
\end{subequations}
and $\bm{n}$ is the unit normal vector pointing outwards from the elements. We note that this form of the numerical flux does not involve an explicit Riemann solver between the left and right states of a given interface. Instead, it is the so-called stabilization matrix $\bm{{{\sigma}}}(\widehat{\bm{u}}_h , \bm{u}_h ; \bm{n})$ that implicitly defines the Riemann solver in hybridized DG methods \cite{Fernandez:AIAA:17a}, and this in turn provides with an implicit subgrid-scale model in large-eddy simulation \cite{Fernandez:JCP_SGS:18c,Fernandez:nonModal:2018}. In this paper, we set $\bm{\sigma} = \lambda_{max} (\widehat{\bm{u}}_h) \, \bm{I}_m$, where $\lambda_{max}$ denotes the maximum-magnitude eigenvalue of $\bm{A}_n = [ \partial \bm{F} / \partial \bm{u} ] \cdot \bm{n}$ and $\bm{I}_m$ is the $m \times m$ identity matrix, and which leads to a Lax-Friedrichs type Riemann solver. 
The hybridized DG discretization of the unsteady compressible Euler equations is obtained by dropping Eq. \eqref{IEDGa} and the viscous terms in Equations \eqref{IEDGb}$-$\eqref{IEDGc}. For additional details on the hybridized DG discretization of the Euler and Navier-Stokes equations, the interested reader is referred to \cite{Fernandez:17a,Fernandez:PhD:2018}.

The semi-discrete system \eqref{IEDG} is further discretized in time using high-order, $L$-stable diagonally implicit Runge-Kutta (DIRK) schemes \cite{Alexander:77}. The use of high-order, $L$-stable methods for the temporal discretization is important for accuracy and robustness when dealing with turbulent shock flows. 
Also, the use of implicit time integration schemes allows to examine the impact of the shock capturing method on the conditioning of the spatial discretization \eqref{IEDG} through the ease of solving the nonlinear system of equations arising from the time discretization. Ill-conditioning of the spatial discretization, which is more difficult to detect with explicit time integration schemes, may yield large degradation errors\footnote{The numerical error is given by the contribution of the projection error $\norm{\bm{u}-\Pi_h(\bm{u})}$ and the degradation error $\norm{\Pi_h(\bm{u})-\bm{u}_h}$, where $\bm{u}$ is the exact solution and $\Pi_h$ the $L^2$ projector onto the approximation space. The projection error is due to the impossibility of representing the exact solution in the approximation space. The degradation error is related to the conditioning of the discrete problem, and increases in general as the discrete problem becomes ill-conditioned.} and deteriorate the accuracy of the numerical solution. 

We emphasize that hybridized DG methods and DIRK methods are considered in this paper for illustration purposes, but the shock capturing method can be used with other spatial and temporal discretization schemes.

\section{\label{s:sensors}Sensors}

In this section, we present physics-based sensors to detect the sharp subgrid-scale features that may appear in the simulation of unsteady turbulent flows, including shock waves and other high-gradient features such as shear and thermal layers.

\subsection*{\label{s:limFun}Limiting function}

\begin{figure}
\centering
\includegraphics[width=0.5\textwidth]{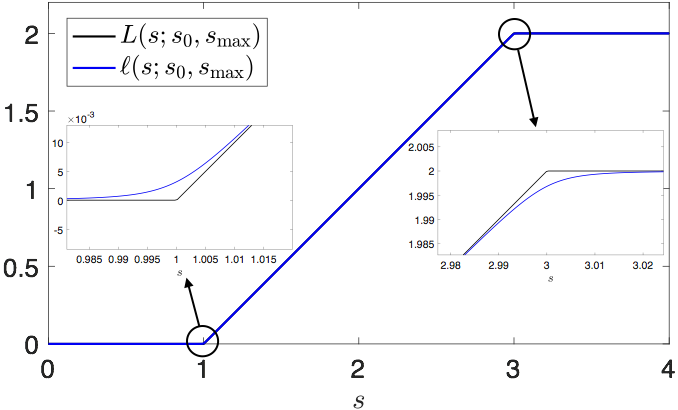}
\caption{An illustrative example of $L(s; s_0, s_{\max})$ and $\ell(s; s_0, s_{\max})$ with $s_0 = 1$ and $s_{\max} = 2$. The inset figures show the close-up view at the two singular points of $L(s; s_0, s_{\max})$.}
\label{limiting}
\end{figure}

It is critical to ensure the sensors remain bounded below by zero and above by an {\it a priori} positive value throughout the simulation, in order to avoid accuracy and stability issues. The lower bound is required to ensure that the artificial viscosities are non-negative, while the upper bound is needed to avoid adding an excessive amount of viscosity. 
Furthermore, it is important to introduce a shift so that the sensors are active only whenever they are above some threshold. 
Suppose that $s_{\min} = 0$ and $s_{\rm max} > 0$ are lower and upper bounds of the sensor $s$. The following function
\begin{equation}
L_0(s; s_{\max}) = \min\{ \max\{s, 0\}, s_{\max}\} ,
\end{equation}
acts as a limiter that strictly enforces the desired property. Shifting the above limiter by $s_0$, we arrive at the following function
\begin{equation}
\label{e:limit}
L(s; s_0, s_{\max}) = \min\{ \max\{s - s_{\rm 0}, 0\}-s_{\rm max}, 0\} + s_{\rm max} , 
\end{equation}
where $s_{\rm 0}$ represents the value of the shift. Since the limiting function \eqref{e:limit} is non-smooth in the sense that its derivative is discontinuous at the points $s_0$ and $s_0 + s_{\rm max}$, it is not well suited for numerical discretization that requires calculation of the partial derivatives. Therefore, we introduce the following smooth approximation
\begin{equation}
\label{smooth}
\ell(s; s_0, s_{\max}) =  \ell_{\min}(\ell_{\max}(s-s_0) - s_{\max}) + s_{\rm max} , 
\end{equation}
where
\begin{equation}
\begin{split}
\ell_{\max}(s) & =  \frac{s}{\pi} \arctan (b s) + \frac{s }{2} - \frac{1}{\pi} \arctan(b) + \frac{1}{2} , \\
\ell_{\min}(s) & = s  - \ell_{\max}(s) , 
\end{split}
\end{equation}
with $b = 100$. Note that $\ell_{\max}(s)$ is a smooth approximation of the max function $\max\{s,0\}$, while $\ell_{\min}(s)$ is a smooth approximation of the min function $\min\{s,0\}$. The function $\ell$ is smooth, and in particular continuously differentiable everywhere. This is important for implicit time integration schemes, in which a nonlinear system of equations needs to be solved at every time step. Discontinuous derivatives can worsen the conditioning of the system and lead to slow convergence or even the crash of the iterative solver. The limiting function $L$ and its smooth approximation $\ell$ for a shift $s_0 = 1$ and an upper bound $s_{\max} = 2$ are illustrated in Figure \ref{limiting}. This particular choice of $s_0$ and $s_{\max}$ will be used for the thermal and shear sensors presented below.

%

\subsection{Shock sensor}


The goal of the shock sensor is to identify shock waves. As such, it is to activate in shocks and vanish elsewhere, including smooth regions of the flow and other sharp features. To this end, we propose a shock sensor $\hat{s}_{\beta}$ of the form
\begin{equation}
\label{e:sbeta}
\hat{s}_{\beta} (\bm{x}) = \ell \big( s_{\beta} ; \ s_{\beta,0} , s_{\beta,\max} \big) , \qquad \qquad s_{\beta} (\bm{x}) = s_{\theta} \cdot s_{\omega} , 
\end{equation}
where $s_{\theta}$ is a dilatation sensor, $s_{\omega}$ is a vorticity (or, more precisely, `anti-vorticity') sensor, and $\bm{x} \in \Omega$ denotes spatial location. The dilatation and vorticity are defined as $\theta = \nabla \cdot \bm{v}$ and $\bm{\omega} = \nabla \times \bm{v}$, respectively. The dilatation sensor is to activate in shock waves. The vorticity sensor is to vanish in vorticity-dominated regions of the flow, in which $| \theta | \ll |\bm{\omega}|$, as well as in non-shocky regions in which $(- \theta)$ is large due to under-resolution. 
In particular, these sensors read as follows
\begin{equation}
s_{\theta} (\bm{x}) = - \frac{h_{\beta}}{k} \frac{\nabla \cdot \bm{v}}{c^*} \ , \qquad \qquad s_{\omega} (\bm{x}) = \frac{(\nabla \cdot \bm{v})^2}{(\nabla \cdot \bm{v})^2 + |\nabla \times \bm{v}|^2 + \epsilon_{\omega}} . 
\end{equation}
The latter sensor was originally proposed by Ducros in \cite{Ducros:1999}. Here,
\begin{equation}
\label{e:h_beta}
h_{\beta} (\bm{x}) = h_{ref} \ \frac{|\nabla \rho|}{\big( \nabla \rho ^t \cdot \bm{M}_h^{-1} \cdot \nabla \rho + \epsilon_h \big)^{1/2}} , 
\end{equation}
is the characteristic element size along the direction of the density gradient, $k$ denotes the polynomial order of the numerical approximation, $c^* = c^*(\bm{x})$ is the critical speed of sound (i.e.\ the speed of sound at sonic conditions), $| \cdot |$ denotes the Euclidean norm, $\bm{M}_h = \bm{M}_h(\bm{x})$ is the metric tensor of the mesh \cite{Li:2005,Yano:2012:PhD}, $h_{ref}$ the reference element size used in the construction of $\bm{M}_h$, and $\epsilon_{\omega}, \epsilon_{h} \sim \epsilon_{m}^2$ are constants of order machine epsilon squared. 
We note that $s_{\beta}$ is uniformly bounded above for stationary $\partial (\cdot) / \partial t = 0$, plane-parallel $\partial (\cdot) / \partial y = \partial (\cdot) / \partial z = 0$ shocks, namely,
\begin{equation}
\label{e:sBetaMaxTh}
\begin{split}
s_{\beta,max}^{th} & = \sup_{M_{1n} \geq 1} s_{\beta} \leq \sup_{M_{1n} \geq 1} s_{\theta} \approx \sup_{M_{1n} \geq 1} \ - \frac{h_{\beta} / k}{c^*} \ \frac{\Delta v_n}{\delta_s} \\ 
&\approx \sup_{M_{1n} \geq 1} \ \frac{h_{\beta} / k}{\delta_s} \ \frac{2 M_{1n}^2 - 2}{(\gamma+1)M_{1n}} \ \sqrt{\frac{\gamma+1}{2 + (\gamma-1) M_{1n}^2}} = \frac{h_{\beta} / k}{\delta_s} \ \frac{2}{\sqrt{\gamma^2-1}} \leq \frac{2}{\sqrt{\gamma^2-1}} , 
\end{split}
\end{equation}
where the superscript $th$ stands for {\it theoretical value}, $M_{1n} = v_{1n} / c_1$ denotes the normal incident Mach number, and $\delta_s$ is the (dynamic) thickness of the shock in the numerical solution. We take advantage of this theoretical upper bound by setting $s_{\beta,max} = 2 / \sqrt{\gamma^2-1}$. Similarly, the use of $s_{\beta,min} = 0$ ensures that no negative artificial bulk viscosity is introduced in the scheme. 
Without these upper and lower bounds in the sensor, numerical oscillations could produce non-physical values of $s_{\beta}$ that positively reinforce the oscillations and lead to nonlinear instability and the simulation breakdown. Finally, $s_{\beta,0}$ is set to $0.01$. This value usually suffices to avoid introducing artificial bulk viscosity away from shocks, while having a minor impact in shock waves.

\subsection{Thermal sensor}

The purpose of the thermal sensor $\hat{s}_{\kappa}$ is to detect thermal gradients that are larger than possible with the grid resolution, and may thus lead to nonlinear instability. In this spirit, we define
\begin{equation}
\label{e:skappa}
\hat{s}_{\kappa} (\bm{x}) = \ell \big( s_{\kappa} ; s_{\kappa,0} , s_{\kappa,max} \big) , \qquad \qquad s_{\kappa} (\bm{x}) = \frac{h_{ref}}{k} \frac{| \nabla_{\bm{\xi}} T |}{T_t} , 
\end{equation}
where $T_t = T_t (\bm{x})$ denotes the stagnation temperature, and $\nabla_{\bm{\xi}} T$ is the temperature gradient under the metric of the reference element, that is,
\begin{equation}
\label{nabla_xi_T}
\nabla_{\bm{\xi}} T = \frac{\partial T}{\partial \bm{\xi}_i} = \sum_{j \leq d} \frac{\partial T}{\partial \bm{x}_j} \frac{\partial \bm{x}_j}{\partial \bm{\xi}_i} = \bm{x}_{\bm{\xi}}^t \cdot \nabla_{\bm{x}} T , \qquad i = 1 , \dots , d . 
\end{equation}
Also, we set $s_{\kappa,0} = 1$ and $s_{\kappa,max} = 2$. The thermal sensor $\hat{s}_{\kappa}$ as a function of $s_{\kappa}$ is plotted in Figure \ref{limiting}. We note that the thermal sensor is active only when $s_{\kappa} > s_{\kappa,0} = 1$, i.e.\ when the thermal gradient cannot be resolved with the mesh resolution. When $s_{\kappa} \leq 1$, i.e.\ when the mesh resolution suffices to resolve the temperature gradient, the thermal sensor is inactive. 
In particular, it can be shown that $s_{\kappa} \lessapprox 4 \gamma / (\gamma+1)^2 \leq 1$ in stabilized stationary normal shock waves, regardless of the incident Mach number. Since shocks will be stabilized by a mechanism that is independent of $\hat{s}_{\kappa}$, namely through artificial bulk viscosity as described below, it is a desired property that the thermal sensor vanishes in shock waves. The upper bound $s_{\kappa,max} = 2$ is used to improve nonlinear stability in a similar way as with the upper bound for the bulk viscosity sensor.



\subsection{Shear sensor}


Like the thermal sensor, the purpose of the shear sensor is to detect velocity gradients that are larger than possible with the grid resolution and may lead to nonlinear instability. In this spirit, we define the shear sensor $\hat{s}_{\mu}$ as
\begin{equation}
\label{smu}
\hat{s}_{\mu} (\bm{x}) = \ell \big( s_{\mu} ; s_{\mu,0} , s_{\mu,max} \big) , \qquad \qquad s_{\mu} (\bm{x}) = \frac{h_{ref}}{k} \ \frac{|| \mathcal{L} (\bm{v}) \cdot \bm{x}_{\bm{\xi}}^t ||_{2} }{v_{max}} , 
\end{equation}
where $|| \cdot ||_{2}$ denotes the spectral norm,
$$ \mathcal{L} (\bm{v}) = \nabla_{\bm{x}} \bm{v} - \textnormal{diag} (\nabla_{\bm{x}} \bm{v}) = \frac{\partial \bm{v}_i}{\partial \bm{x}_j} \Big( 1 - \delta_{ij} \Big), $$
and
$$ v_{max} (\bm{x}) = \bigg( \norm{\bm{v}}^2 + \frac{2}{\gamma - 1} \ c^2 \bigg) ^{1/2} $$
is the {\it maximum isentropic velocity}, defined as the velocity the flow if all total energy was converted into kinetic energy through an isentropic expansion. The presence of the $\textnormal{diag}(\nabla_{\bm{x}} \bm{v})$ term in $\mathcal{L} (\bm{v})$ is for the shear sensor to vanish in shock waves, as these will be stabilized through artificial bulk viscosity instead.

For the same reasons as with the thermal sensor, we choose $s_{\mu,0} = 1$ and $s_{\mu,max} = 2$. The latter improves nonlinear stability and the former ensures the sensor activates only for sharp shear features that may potentially lead to numerical instability.

\section{\label{s:AV}Stabilization through artificial viscosities}


We increase selected fluid viscosities to resolve sharp features over the smallest distance allowed by the grid resolution. The bulk viscosity, thermal conductivity and shear viscosity are thus given by the contribution of the physical $(\beta_f, \kappa_f, \mu_f)$ and artificial $(\beta^*, \kappa^*, \mu^*)$ values, that is,
$$ \beta = \beta_f + \beta^* , \qquad \qquad \kappa = \kappa_f + \kappa^* = \kappa_f + \kappa_1^* + \kappa_2^* , \qquad \qquad \mu = \mu_f + \mu^* . $$
Shock waves, thermal gradients, and shear layers are stabilized by increasing the bulk viscosity, thermal conductivity, and shear viscosity, respectively. Contact discontinuities are stabilized through one or several of these mechanisms, depending on their particular structure. The thermal conductivity is also augmented in hypersonic shock waves through the term $\kappa_1^*$, as discussed below. Our stabilization procedure is consistent with mathematical and physical arguments that identify these as the mechanisms responsible for stabilizing 
these various flow features. Also, it is consistent with our choice of sensors in the sense that the penalty is imposed directly on the quantities that are used for sensing.

We emphasize that shock waves are stabilized through $\beta^*$ and $\kappa_1^*$ only. The latter term is used in hypersonic shocks only, in order to improve nonlinear stability and make the thermal thickness of the shock $\theta_s$ of the same order as the dynamic thickness $\delta_s$. Note that $\theta_s \approx \delta_s$ is obtained in non-hypersonic shocks even without $\kappa_1^*$. Also, while artificial shear viscosity can also stabilize shock waves, the bulk viscosity has a much smaller impact on the dissipation of vortical structures crossing the shock and is thus preferred for LES applications. 
Finally, we note that the jump conditions across a shock wave (i.e.\ the Rankine-Hugoniot conditions) are given by conservation arguments on a larger scale than the shock wave thickness and do not depend on the constitutive equation for the viscous stresses inside the shock wave (as these appear inside of the divergence operator and their contribution vanishes on scales that are larger than the shock wave thickness). 
Thus, the jump conditions are not violated by the use of artificial physical viscosities inside the shock wave.

The artificial viscosities are devised such that the cell P\'eclet number 
is of order $1$ (note the sensors are of order $1$ when active), and in particular are given by
\begin{subequations}
\label{artTransCoef}
\begin{align}
\beta^{*}(\bm{x}) &= \Phi_{\beta} \bigg[ \rho \ \frac{k_{\beta} \ h_{\beta}}{k} \ \big( \norm{\bm{v}}^2 + c^{*2} \big) ^{1/2} \ \hat{s}_{\beta} \bigg] , \\
\kappa^*(\bm{x}) &= \kappa_1^* + \kappa_2^* = \Phi_{\beta} \bigg[ \frac{c_p}{Pr_{\beta}^*} \ \bigg( \rho \ \frac{k_{\beta} \ h_{\beta}}{k} \ \big( \norm{\bm{v}}^2 + c^{*2} \big) ^{1/2} \ \hat{s}_{\beta} \bigg) \bigg] + \Phi_{\kappa} \bigg[ \rho \ c_p \ \frac{k_{\kappa} \ h_{\kappa}}{k} \ \big( \norm{\bm{v}}^2 + c^{*2} \big) ^{1/2} \ \hat{s}_{\kappa} \bigg] , \\
\mu^*(\bm{x}) &= \Phi_{\mu} \bigg[ \rho \ \frac{k_{\mu} \ h_{\mu}}{k} \ \big( \norm{\bm{v}}^2 + c^{*2} \big) ^{1/2} \ \hat{s}_{\mu} \bigg] .
\end{align}
\end{subequations}
Here $\Phi_{ \{ \beta, \kappa, \mu \} } \big[ \cdot \big]$ are smoothing operators, $Pr_{\beta}^*$ is an artificial Prandtl number relating $\beta^*$ and $\kappa_1^*$, $k_{ \{ \beta , \kappa , \mu \} }$ are positive constants, and
\begin{subequations}
\label{e:hs}
\begin{align}
h_{\kappa} (\bm{x}) & = h_{ref} \ \frac{|\nabla_{\bm{x}} T|}{\big( \nabla_{\bm{x}} T ^t \cdot \bm{M}_h^{-1} \cdot \nabla_{\bm{x}} T + \epsilon_h \big)^{1/2}} , \\
h_{\mu} (\bm{x}) & = h_{ref} \ \sigma_{min} (\bm{M}_h) = h_{ref} \ \inf_{|\bm{a}| = 1} \big\{ \bm{a}^t \cdot \bm{M}_h \cdot \bm{a} \big\} , 
\end{align}
\end{subequations}
are the element size in the direction of the temperature gradient and the smallest element size among all possible directions, respectively. The remaining nomenclature in Equations \eqref{artTransCoef}$-$\eqref{e:hs} was introduced in the previous sections. 
Theoretical estimates of $k_{\beta}$ and $Pr_{\beta}^*$ to optimally resolve a stationary normal shock over a thickness $\delta_s , \theta_s \approx h_{\beta} / k$ are presented in \ref{s:optimalKkappa}. In particular, we set
\begin{equation}
\label{e:PrBetaStare2}
k_{\beta} = 1.5 , \qquad \qquad Pr_{\beta}^*(\bm{x}) = Pr_{\beta,min}^* \, \Big[ 1 + \exp \big( - 2 \, \alpha_{Pr_{\beta}^*} \, (M - M_{thr} ) \big) \Big] , 
\end{equation}
where $M = M(\bm{x})$ denotes the local Mach number, $M_{thr} = 3$ is a threshold Mach number, $Pr_{\beta,min}^* = 0.9$, and $\alpha_{Pr_{\beta}^*} = 2$. These constants have been tuned to obtain sharp, non-oscillatory one-dimensional shocks over a wide range of Mach numbers while using the smallest possible amount of artificial viscosity. Note that $Pr_{\beta}^* = 0.9$ would also lead to well-resolved shocks at the expense of introducing unnecessary thermal conductivity in non-hypersonic shocks. Also, the local Mach number in Eq. \eqref{e:PrBetaStare2} can be replaced by a (constant) reference Mach number, such as the freestream Mach number in the case of external flows. This simpler choice makes the $Pr_{\beta}^*$ field constant in the computational domain but could negatively impact the weak shock waves that may spontaneously appear in highly compressible turbulent flows. 
Finally, we set $k_{ \{ \kappa , \mu \} } = 1.0$.

\subsection*{\label{s:smoothingOperator}Smoothing operators}

Large inter-element jumps in the numerical solution, as it occurs in under-resolved sharp features, lead to large discontinuouties in the artificial viscosity fields. According the previous results in the literature \cite{Barter:2010,Persson:2013} and our own experience, this may degrade the accuracy of the solution and lead to numerical stability issues. Hence, we equip the artificial viscosities with smoothing operators $\Phi_{ \{ \beta, \kappa, \mu \} }$ that map onto a $\mathcal{C}^{\alpha}$-continuous space ($\alpha \geq 0$). In our experience, further smoothness beyond $\alpha = 0$ does not provide additional stability. This is consistent with the fact that the artificial viscosities only enter in the discrete system \eqref{IEDG} through the terms $( \bm{F} + \bm{G} , \nabla \bm{w} ) _{\mathcal{T}_h}$, $\langle \widehat{\bm{f}}_h + \widehat{\bm{g}}_h, \bm{w} \rangle_{\partial \mathcal{T}_h}$, $\langle \widehat{\bm{f}}_h + \widehat{\bm{g}}_h, \bm{\mu} \rangle_{\partial \mathcal{T}_h \backslash \partial \Omega}$ and $\langle \widehat{\bm{b}}_h, \bm{\mu} \rangle_{\partial \Omega}$, and the notion of $\mathcal{C}^{\alpha}$-continuity for $\alpha > 0$ is thus lost upon discretization. (Indeed, even a weaker condition that continuity would suffice for the purpose of the smoothing operator.) 
Also, positivity of $\Phi_{ \{ \beta, \kappa, \mu \} }$, in the functional analysis sense, is important to ensure the artificial viscosities are pointwise non-negative.

Convolution with a truncated Gaussian filter \cite{Cook:2005}, projection onto a lower dimensional continuous approximation space, and elementwise reconstruction procedures \cite{Moro:2016} are examples of smoothing operators. The appropiate choice of smoothing operator depends on the type and accuracy order of the numerical scheme. In this paper, we employ an elementwise linear reconstruction procedure analogous to that introduced in \cite{Moro:2016} for the element size. Devising new smoothing operators is beyond the scope of this work.

\subsection*{Other comments and practical suggestions}
\begin{itemize}
\item In the context of implicit time integration schemes, the artificial viscosities can be computed using the solution at the end of the previous time step (or, in the case of multi-stage methods, at the end of the previous time stage) or using the solution at the end of the current time step (or time stage in multi-stage methods). For the small time-step sizes required in large-eddy simulation, no significant differences have been observed between both approaches. The former one is adopted for the numerical results in Section \ref{s:results}.
\item If the latter approach was used, negative thermodynamic quantities, such as negative density and pressure, could be potentially encountered during the iterative procedure used to solve the system of equations arising from the discretization. For this reason, we suggest replacing the thermodynamic quantities involved in the evaluation of the sensors and artificial viscosities by smooth strictly positive surrogates, e.g.\ by limiting functions similar to those introduced in Section \ref{s:limFun}.
\item For the simulation of inviscid flows, we suggest suppressing the artificial viscosities near slip walls in order to ensure well-posedness of the discretization.
\end{itemize}

\section{\label{s:results}Numerical results}

We examine the performance of the shock capturing method for unsteady flows in transonic, supersonic and hypersonic regimes. The robustness, shock resolution and impact of the model on the turbulent structures and acoustic waves are investigated. Two-dimensional and three-dimensional problems, as well as different accuracy orders, are considered. All results are presented in non-dimensional form. $Pr_f = c_p \, \mu_f / \kappa_f = 0.71$, $\beta_f = 0$ and $\gamma = 1.4$ are assumed in all the test problems.

%
%
%

\subsection{Inviscid strong-vortex/shock-wave interaction}

\subsubsection{Case description and numerical discretization}

We consider the two-dimensional inviscid interaction between a strong vortex and a shock wave. The problem domain is $\Omega = (0 , 2 L) \times (0 , L)$ and a stationary normal shock wave is located at $x_s = L / 2$. A counter-clockwise rotating vortex is initially located upstream of the shock and advected downstream by the inflow velocity. In particular, the inflow Mach number is $M_{\infty} = 1.5$ and the vortex is initially radius $b = 0.175 \, L$ and centered at $(x, y) = (L / 4, \, L / 2)$. The top and bottom boundaries are slip adiabatic walls. The initial velocity, temperature, density and pressure fields upstream the shock are given by
\begin{subequations}
\begin{equation}
\bm{v} (r) = \bm{v}_{\theta} (r) + u_{\infty} \ \hat{\bm{e}}_x , \qquad \qquad \bm{v}_{\theta} (r) = u_m \ \hat{\bm{e}}_{\theta} \cdot \begin{cases} 
      \frac{r}{a} & \text{if } r \leq a , \\
      \frac{a}{a^2-b^2} \Big( r - \frac{b^2}{r} \Big) & \text{if } a \leq r \leq b , \\
      0 & \text{if } b \leq r , 
   \end{cases}
\end{equation}
\begin{equation}
T(r) = \begin{cases} 
      T_{\infty} - \int_r^b \frac{1}{c_p} \frac{|\bm{v}_{\theta}(r')|^2}{r'} \, dr' & \text{if } r < b , \\
      T_{\infty} & \text{if } b \leq r , \\
   \end{cases} , \quad \quad \rho \, (r) = \rho_{\infty} \bigg( \frac{T(r)}{T_{\infty}} \bigg) ^{\frac{1}{\gamma - 1}} , \quad \quad p \, (r) = p_{\infty} \bigg( \frac{T(r)}{T_{\infty}} \bigg) ^{\frac{\gamma}{\gamma - 1}} , 
\end{equation}
\end{subequations}
where $a = 0.075 \, L$ is a constant, $r$ denotes the distance to the vortex center, $u_m = 3 \, u_{\infty}\, / \, 5$ is the maximum tangential velocity of the vortex, $u_{\infty}$ the inflow velocity magnitude,  $T_{\infty}$ the inflow temperature, and $\hat{\bm{e}}_x$ and $\hat{\bm{e}}_{\theta}$ are unit vectors along the $x$- and the tangential (around the vortex center) directions, respectively.
The initial condition downstream the shock wave is given by one-dimensional stationary shock wave theory. This completes the non-dimensional description of the problem. While not turbulent, this unsteady laminar flow serves as a preliminary test case to examine the performance of the shock capturing method.


The problem domain is partitioned into $400 \times 200$ uniform quadrilateral elements, and the time-step size is set to $\Delta t = 3.00 \cdot 10^{-4} \, L \, u_{\infty}^{-1}$. Sixth-order IEDG and third-order DIRK(3,3) schemes are used for the spatial and temporal discretization, respectively. Slip, adiabatic wall boundary conditions are imposed on the top and bottom surfaces, whereas the characteristics-based, non-reflecting boundary condition in \cite{Fernandez:17a} is used on the inflow and outflow.

 \begin{figure}
 \centering
 {\includegraphics[width=1.0\textwidth]{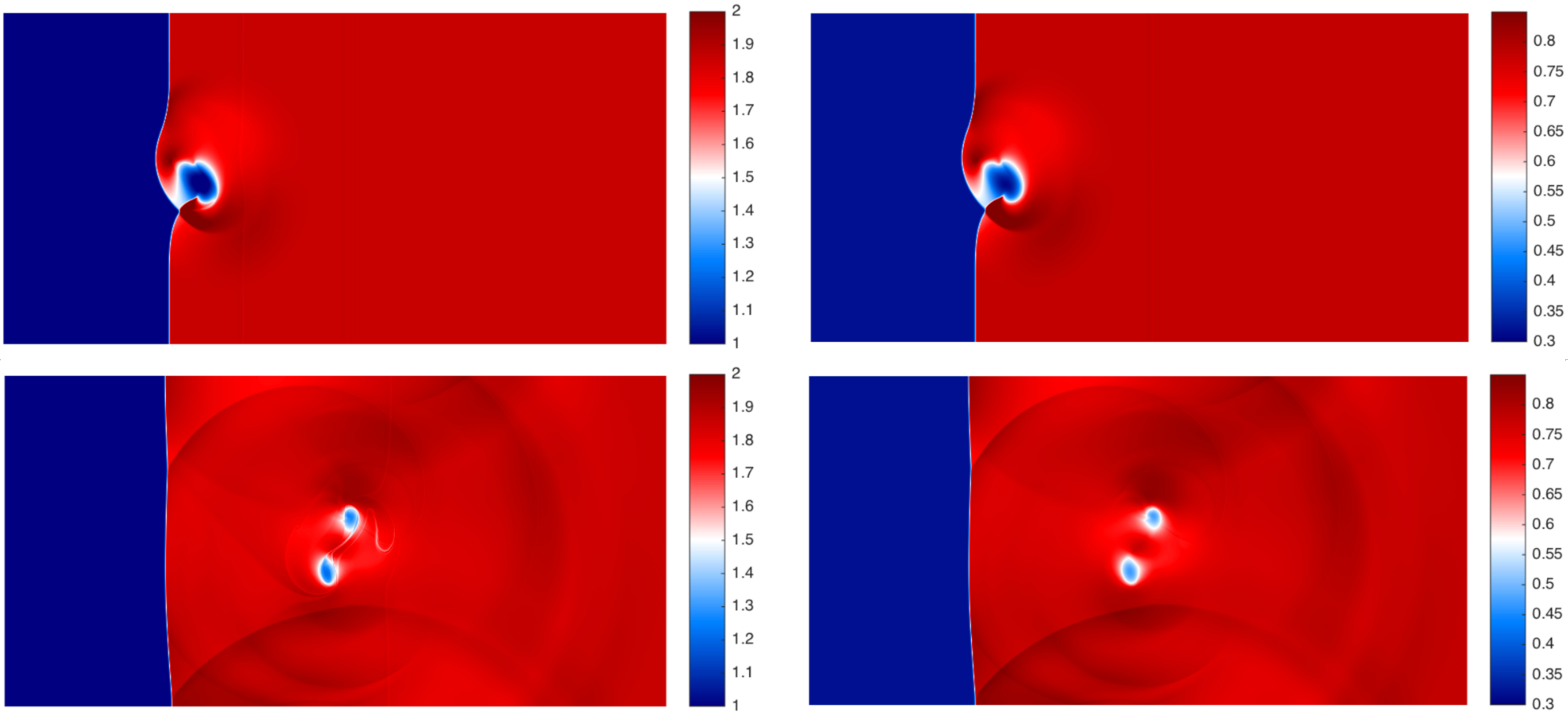}}
  \caption{Non-dimensional density $\rho / \rho_{\infty}$ (left) and pressure $p / (\rho_{\infty} u_{\infty}^2)$ (right) fields of the strong-vortex/shock-wave interaction problem at times $t_1 = 0.35 \, \gamma^{1/2} L \, u_{\infty}^{-1}$ (top) and $t_2 = 1.05 \, \gamma^{1/2} L \, u_{\infty}^{-1}$ (bottom). After the shock wave and the vortex meet, strong acoustic waves are generated and propagate on the downstream side of the shock.}\label{f:densityPressureFields_VSI}
 \end{figure}

  \begin{figure}
 \centering
 {\includegraphics[width=0.8\textwidth]{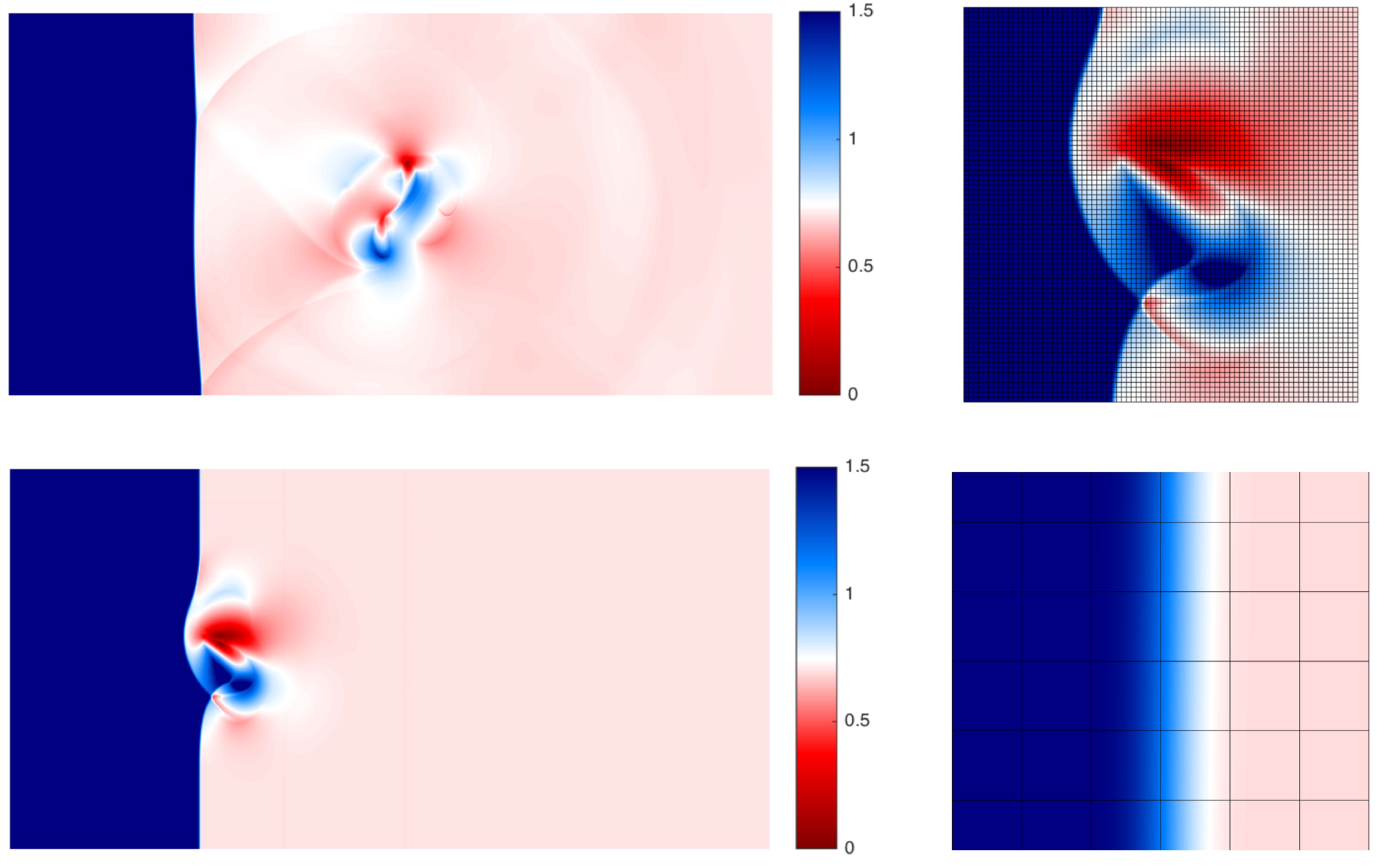}}
 \caption{Mach number field of the strong-vortex/shock-wave interaction problem at times $t_1 = 0.35 \, \gamma^{1/2} L \, u_{\infty}^{-1}$ (top) and $t_2 = 1.05 \, \gamma^{1/2} L \, u_{\infty}^{-1}$ (bottom). Zooms around the shock wave are shown on the right images. The shock is non-oscillatory and resolved within one element.}\label{f:machField_VSI}
 \end{figure} 

 \begin{figure}
 \centering
 {\includegraphics[width=0.32\textwidth]{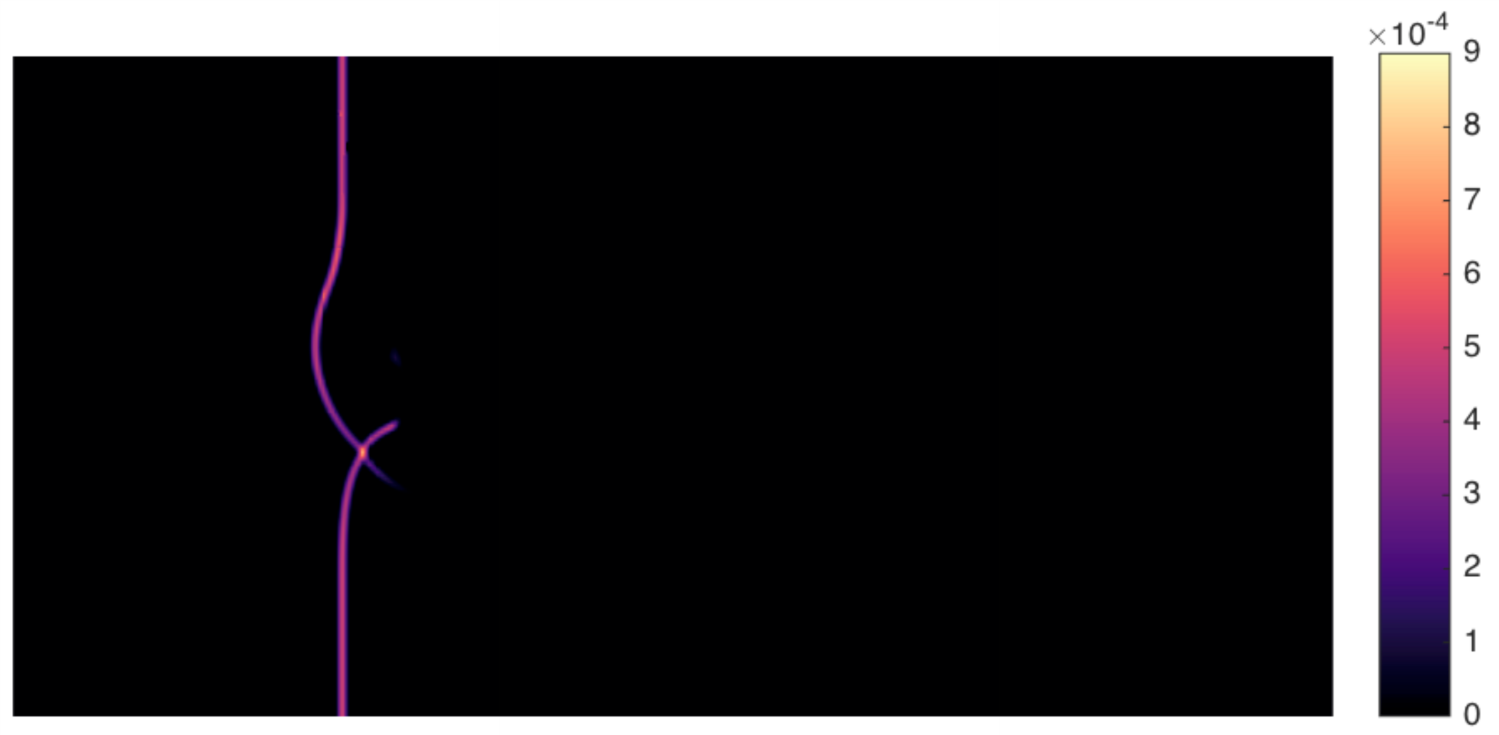}}
 \hfill {\includegraphics[width=0.32\textwidth]{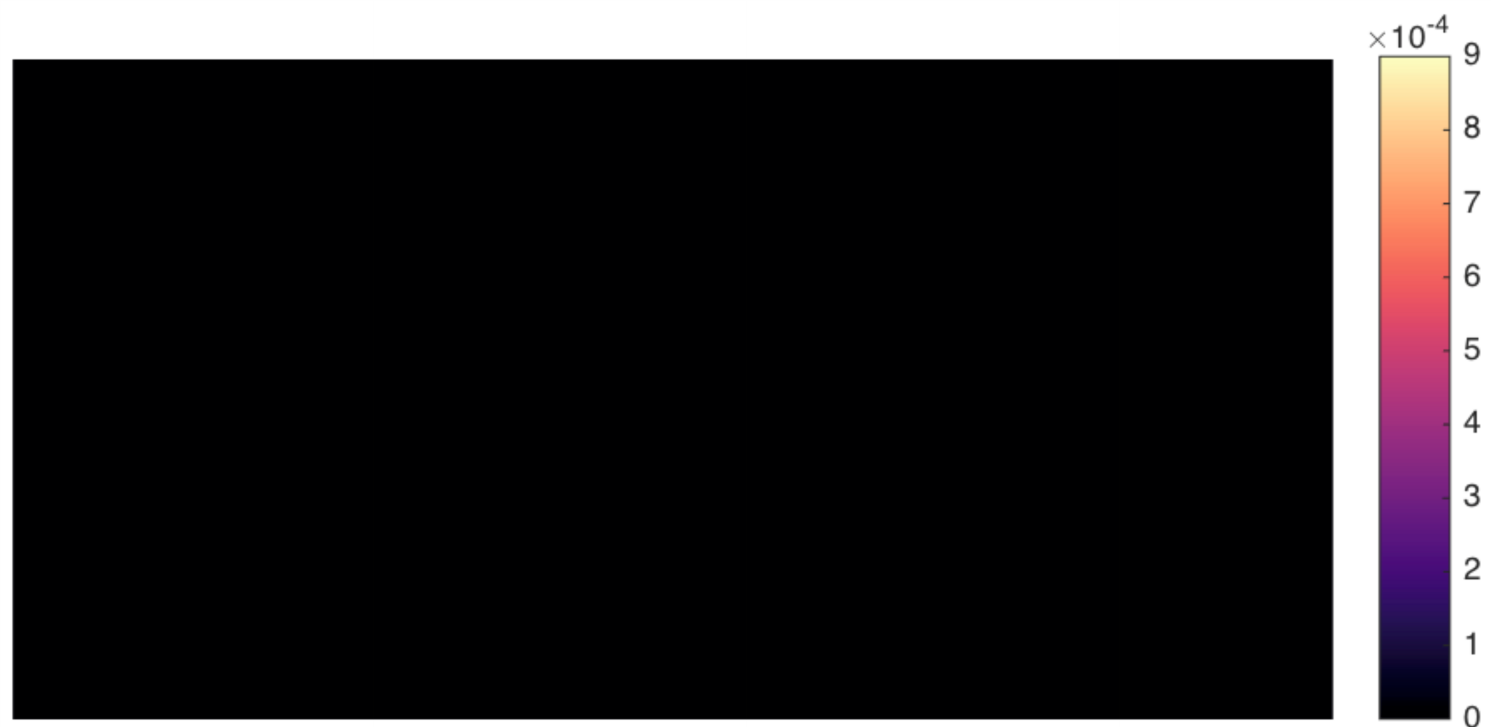}}
 \hfill {\includegraphics[width=0.32\textwidth]{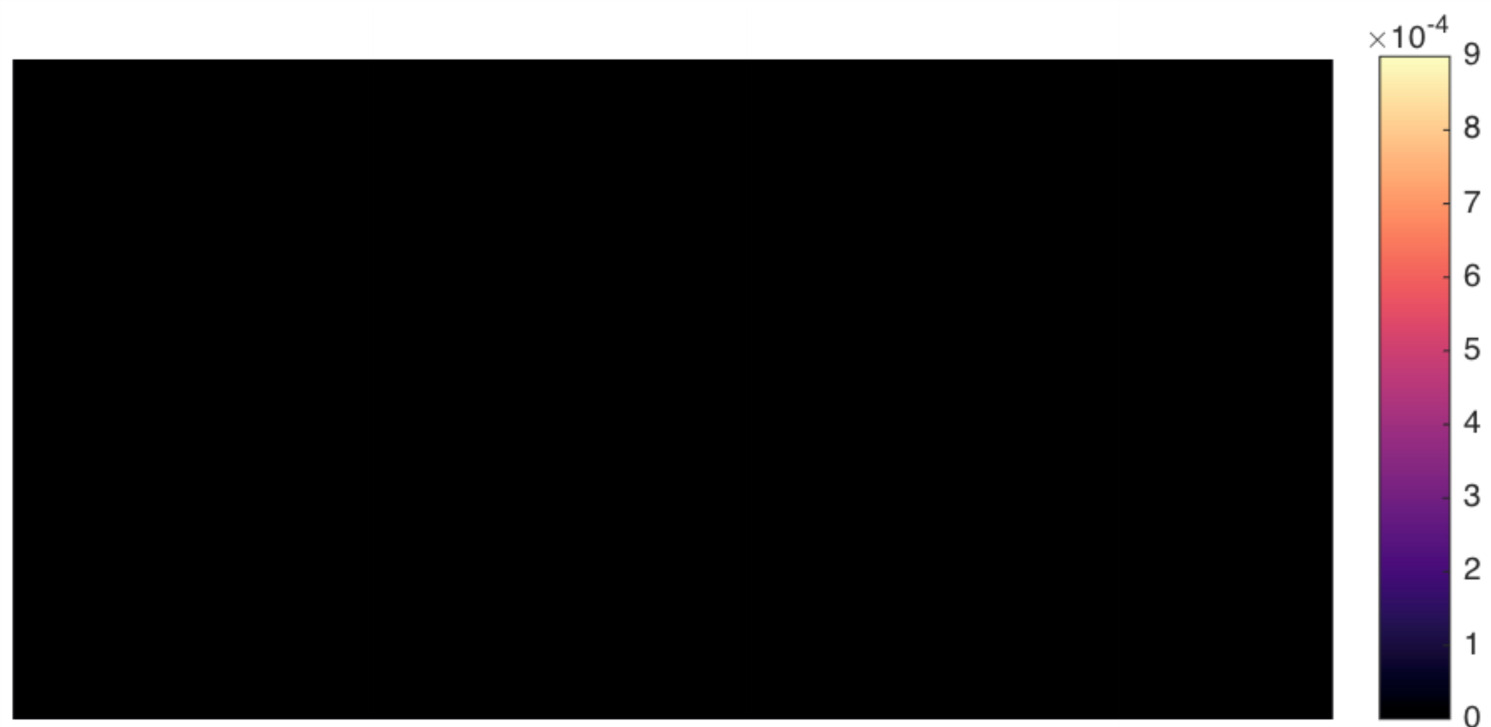}}
 \hfill {\includegraphics[width=0.32\textwidth]{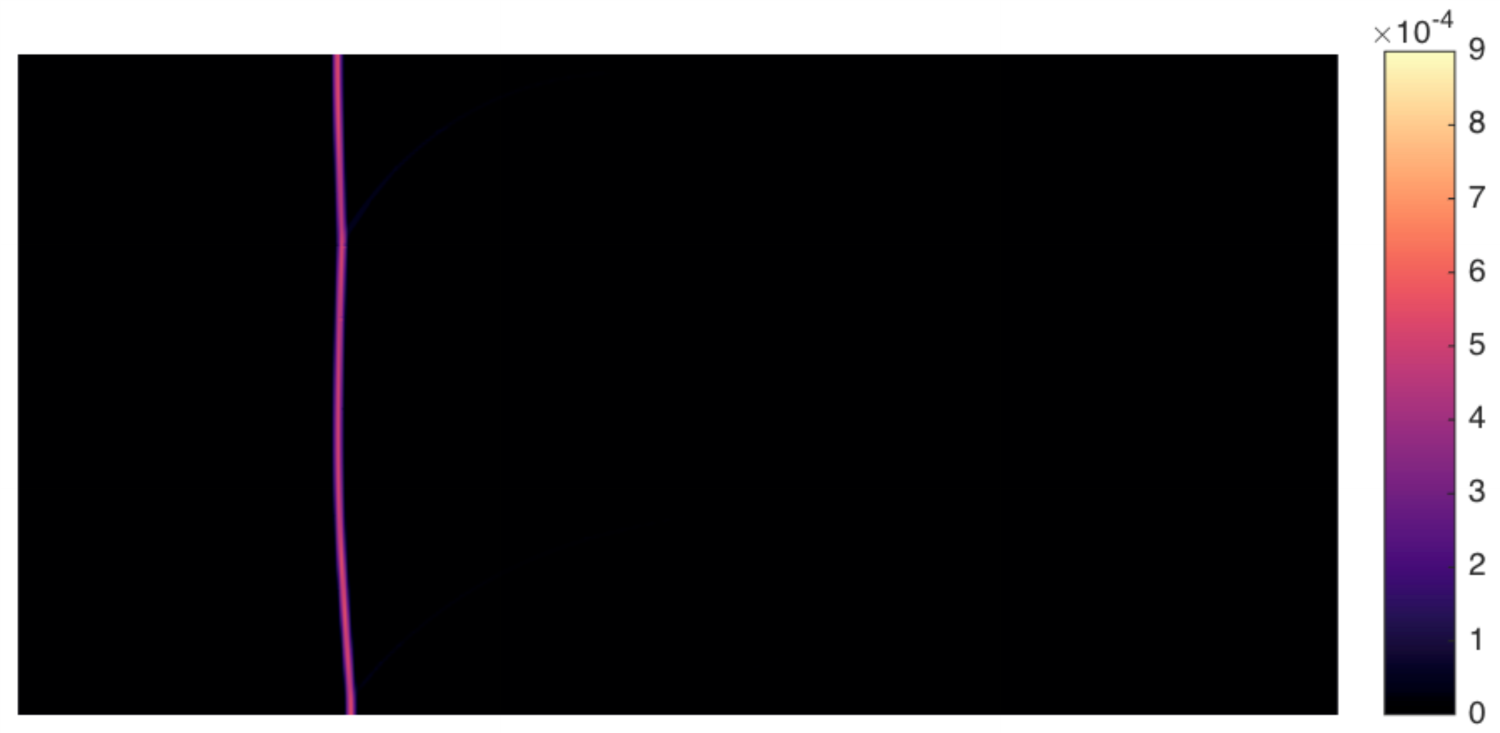}}
 \hfill {\includegraphics[width=0.32\textwidth]{kappaField_VSI_t2.pdf}}
 \hfill {\includegraphics[width=0.32\textwidth]{muField_VSI_t2.pdf}}
  \caption{Snapshot of the non-dimensional artificial bulk viscosity $\beta^* \rho_{\infty}^{-1} u_{\infty}^{-1} L^{-1}$ (left), artificial thermal conductivity $\kappa_2^* \, \rho_{\infty}^{-1} u_{\infty}^{-1} c_v^{-1} L^{-1}$ (center) and artificial shear viscosity $\mu^* \rho_{\infty}^{-1} u_{\infty}^{-1} L^{-1}$ (right) fields of the strong-vortex/shock-wave interaction problem at times $t_1 = 0.35 \, \gamma^{1/2} L \, u_{\infty}^{-1}$ (top) and $t_2 = 1.05 \, \gamma^{1/2} L \, u_{\infty}^{-1}$ (bottom). Note $\beta^*$ vanishes outside the shock, including the strong vortex and acoustic waves.}\label{f:AV_fields_VSI}
 \end{figure}

\subsubsection{Numerical results}

Figure \ref{f:densityPressureFields_VSI} shows the density and pressure fields at the times $t_1 = 0.35 \, \gamma^{1/2} L \, u_{\infty}^{-1}$ and $t_2 = 1.05 \, \gamma^{1/2} L \, u_{\infty}^{-1}$. When the shock wave and the vortex meet, the former is distorted and the latter split into two separate vortical structures. Strong acoustic waves are then generated from the moving vortex and propagate on the downstream side of the shock. The Mach number fields, together with zooms around the shock wave and the details of the computational mesh, are shown in Figure \ref{f:machField_VSI}. The shock is non-oscillatory and resolved within one element.


The artificial bulk viscosity $\beta^*$, artificial thermal conductivity $\kappa_2^*$ and artificial shear viscosity $\mu^*$ fields at the target times are shown in Figure \ref{f:AV_fields_VSI}. Despite the strong pressure waves and the correspondingly large negative velocity divergence at $t_2$, the artificial bulk viscosity vanishes everywhere outside the shock wave. Similarly, it is active only in the shock wave at $t_1$ despite the strong interaction between the vortex and the shock at this time. 
Note there are no other sharp features than the shock wave in this problem and thus $\kappa_2^*$ and $\mu^*$ vanish in the entire domain.

\subsection{\label{s:hypersonicCylinder}Two-dimensional hypersonic cylinder}

\subsubsection{Case description and numerical discretization}

The second numerical example is the hypersonic flow around a two-dimensional adiabatic cylinder at Reynolds number $Re_{\infty} = \rho_{\infty} \, u_{\infty} \, d / \mu = 376,930$ and Mach number $M_{\infty} = u_{\infty} / c_{\infty} = 17.605$, where $\rho_{\infty}$, $u_{\infty}$, $c_{\infty}$, and $d$ denote the freestream density, freestream velocity, freestream speed of sound, and cylinder diameter, respectively. The computational domain spans 2.5 diameters away from the center of the cylinder and is discretized using an isoparametric O-mesh with 16,000 quadrilateral elements. The time-step size is set to $\Delta t = 10^{-3} \, d / u_{\infty}$. Forth-order HDG and third-order DIRK(3,3) schemes are used in this example.




\subsubsection{Numerical results}

\begin{figure}
\centering
\includegraphics[width=0.9\textwidth]{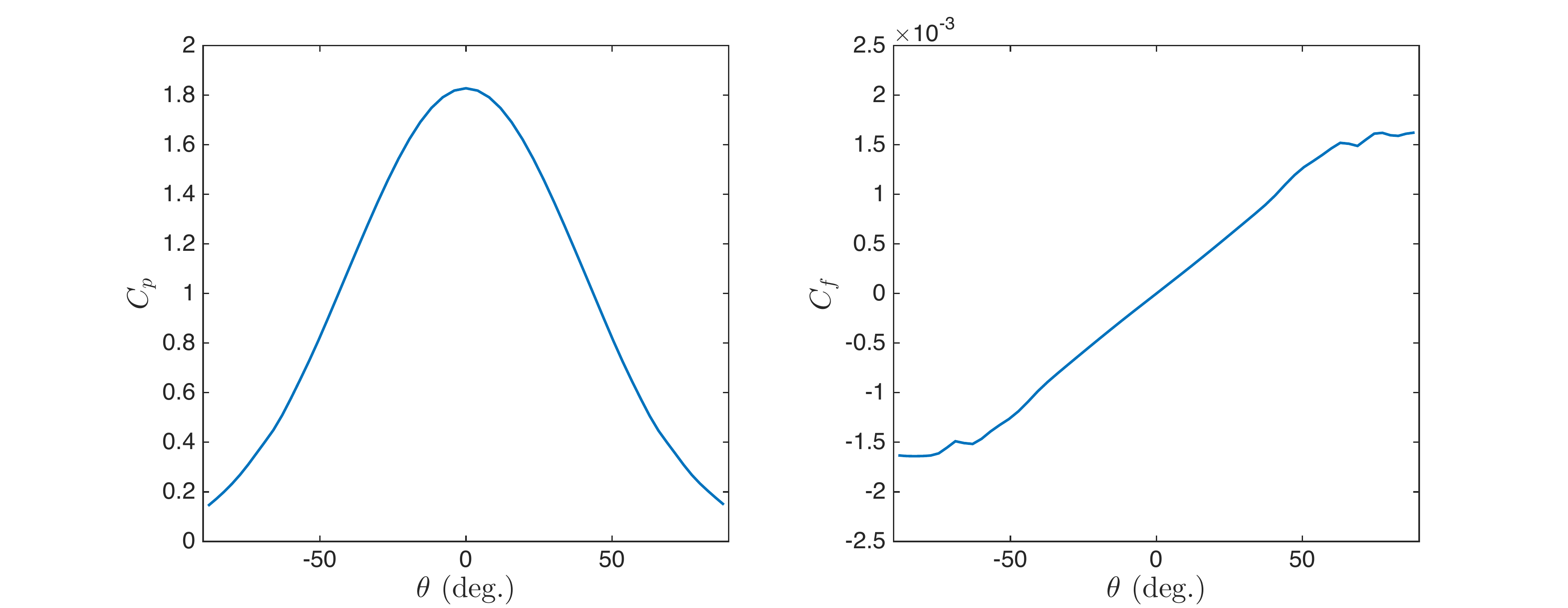}
\caption{\label{f:Cp_Cf_Cyl} Time-averaged pressure (left) and skin friction (right) coefficients on the upstream half of the hypersonic cylinder.}
\end{figure}

\begin{figure}
 \centering
 {\includegraphics[width=0.49\textwidth]{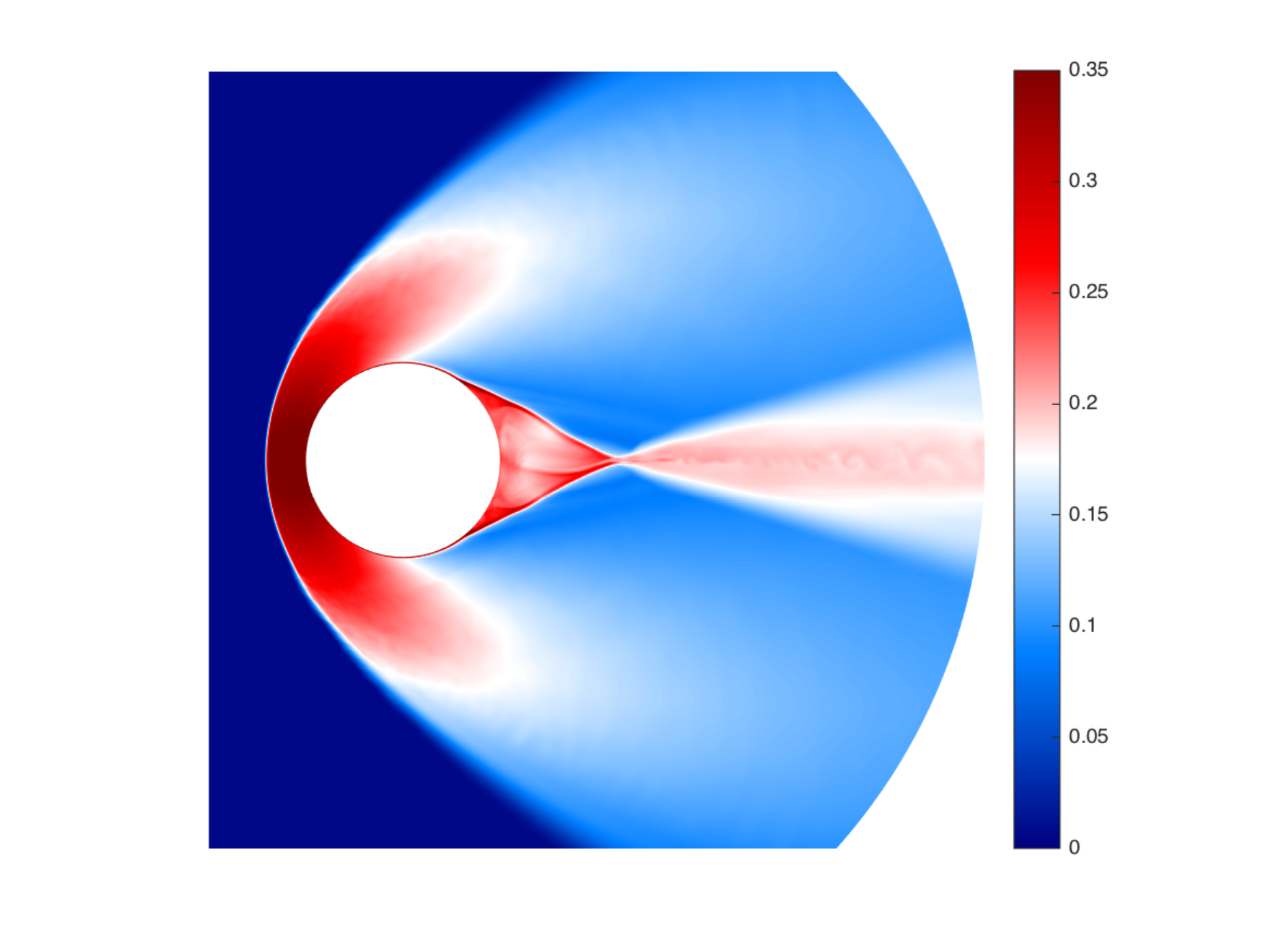}}
 \hfill {\includegraphics[width=0.49\textwidth]{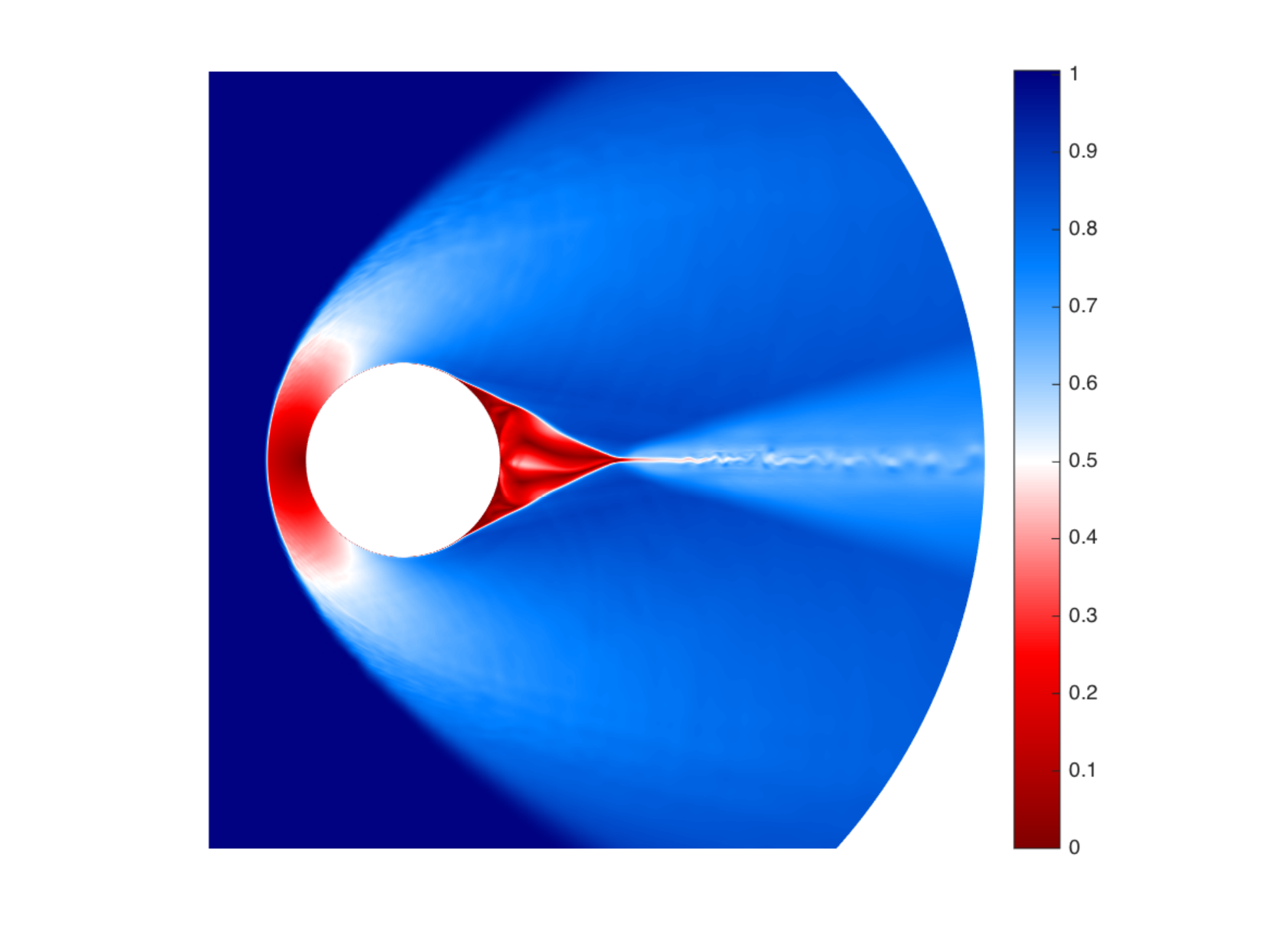}}
 \hfill {\includegraphics[width=0.49\textwidth]{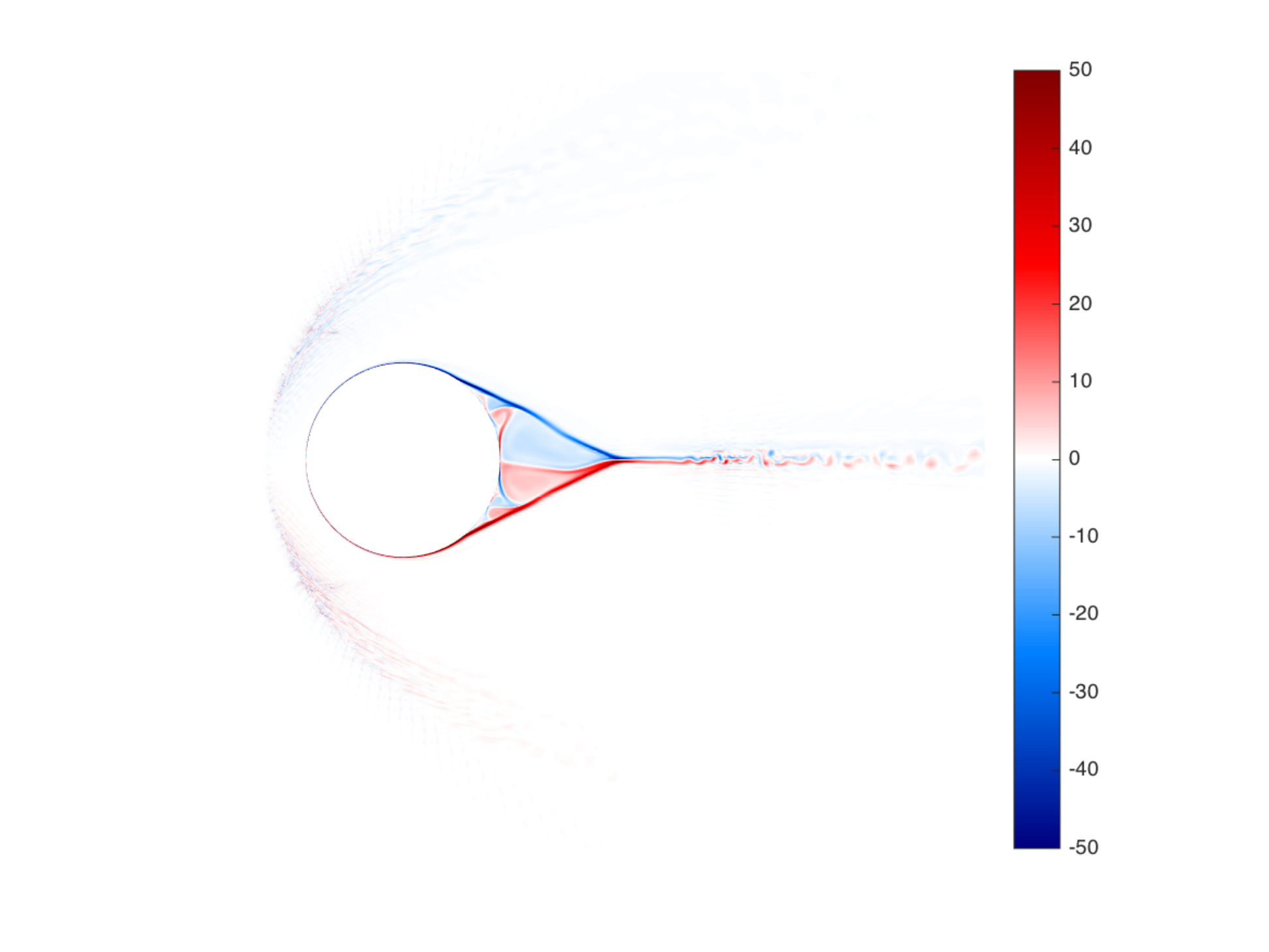}}
 \hfill {\includegraphics[width=0.49\textwidth]{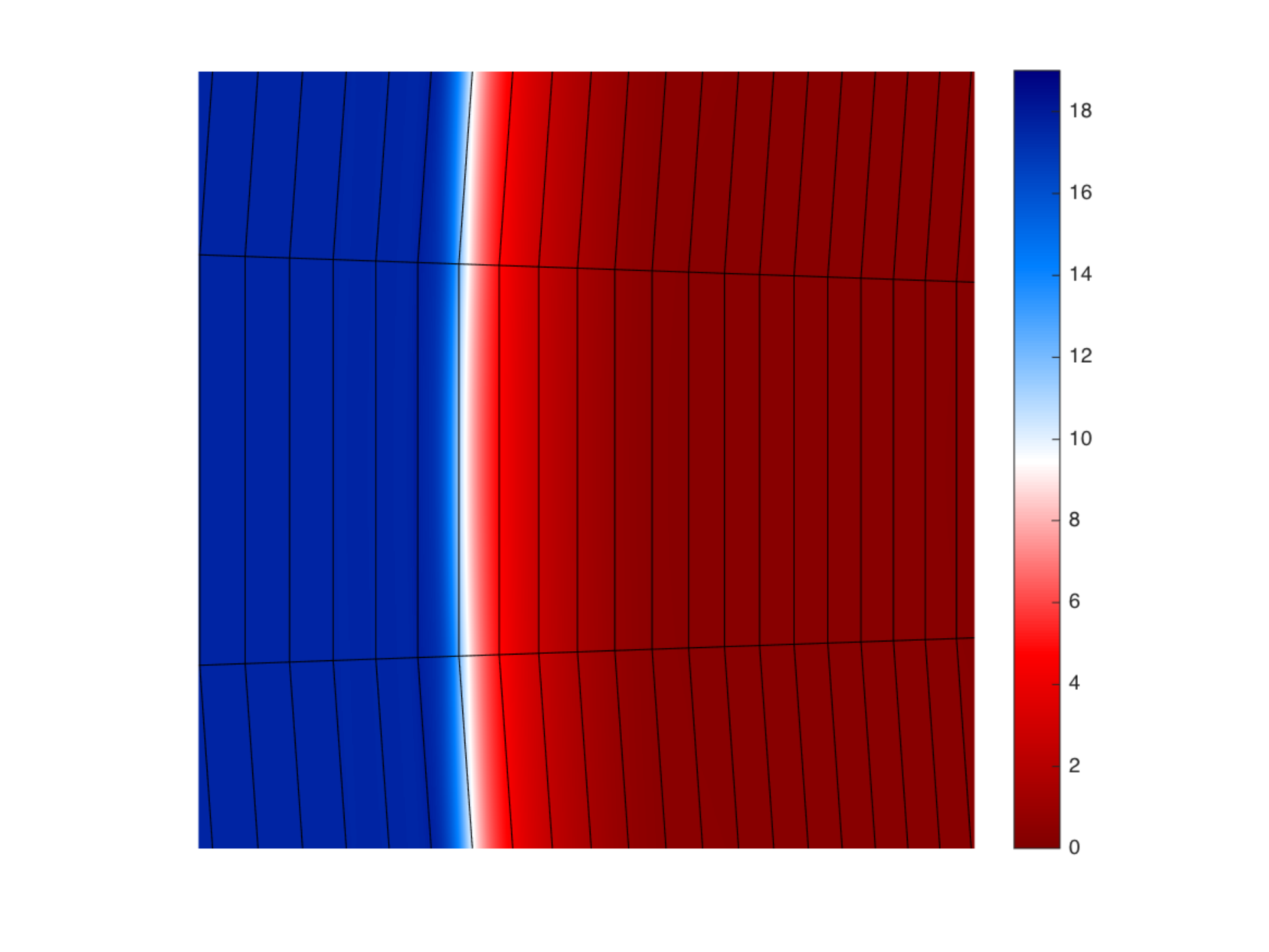}}
  \caption{Snapshot of the non-dimensional temperature $c_v \, T / u_{\infty}^2$ (top left), velocity magnitude $|\bm{v}| / u_{\infty}$ (top right) and vorticity $d \, \omega / u_{\infty}$ (bottom left) fields for the hypersonic cylinder. A zoom of the Mach number field around the center of the shock is shown in the bottom right image. The shock is non-oscillatory and resolved within three elements.}\label{f:tempVelMagVortCylinder}
 \end{figure}

Figure \ref{f:Cp_Cf_Cyl} shows the time-averaged pressure (left) and skin friction (right) coefficients on the upstream half of the cylinder. Snapshots of the temperature, velocity magnitude and vorticity fields are shown in Figure \ref{f:tempVelMagVortCylinder}. A zoom of the Mach number field around the center of the shock, together with the computational mesh, are shown in the bottom right of this figure. Despite the high incident Mach number, the shock is non-oscillatory and resolved within three elements.

\begin{figure}
 \centering
 \hfill {\includegraphics[width=0.49\textwidth]{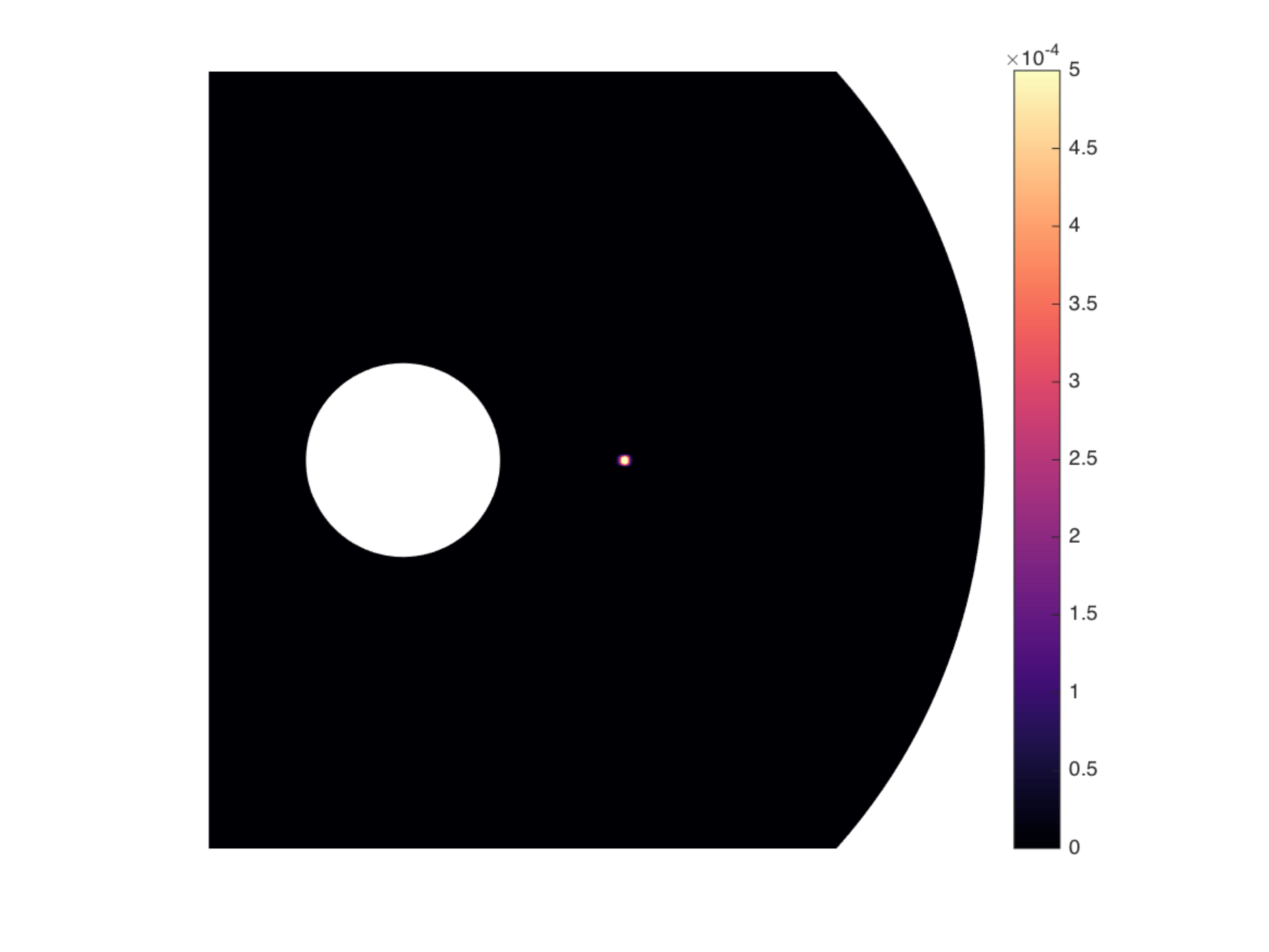}}
 \hfill {\includegraphics[width=0.49\textwidth]{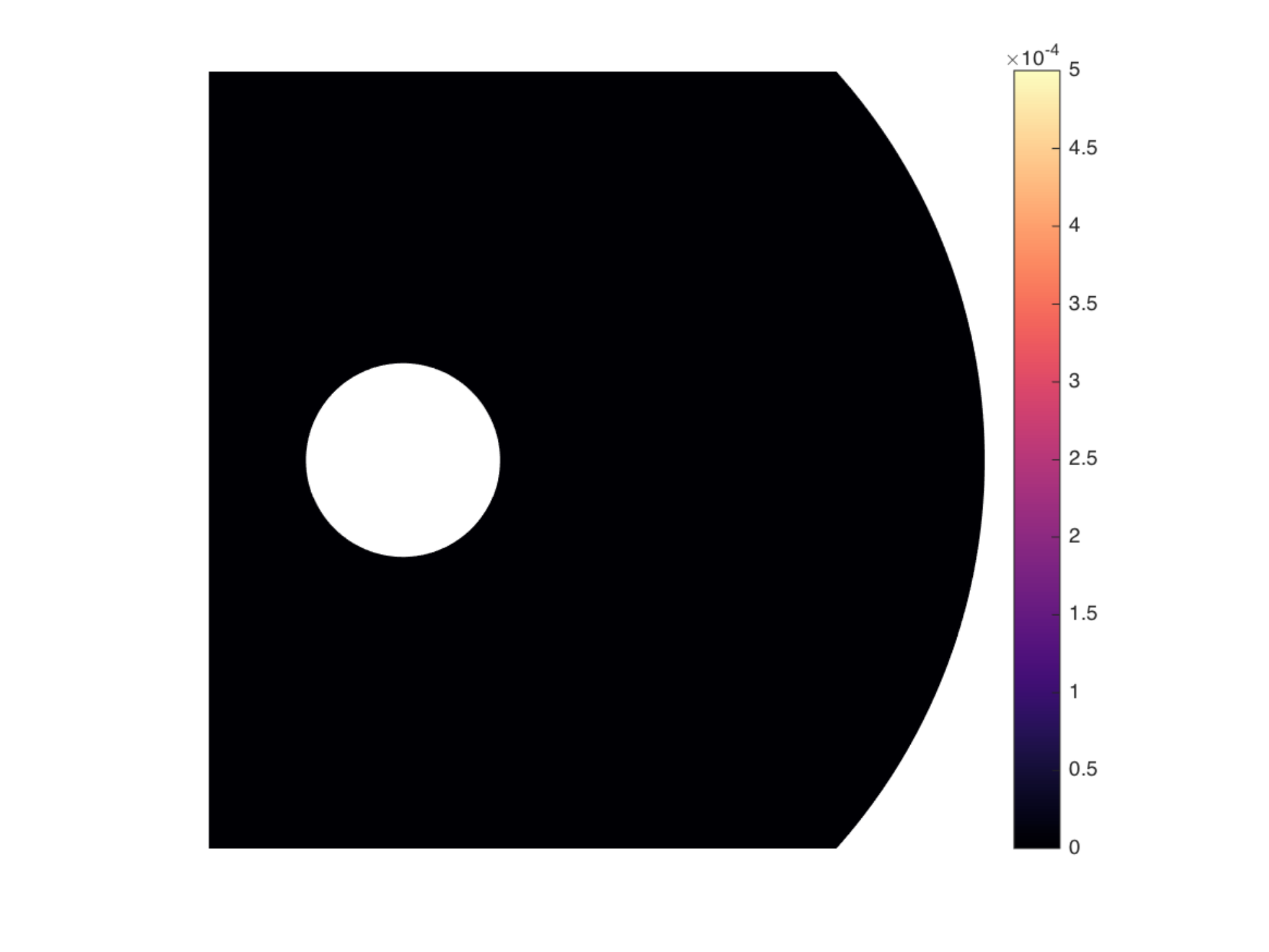}}
  \caption{Snapshot of the non-dimensional artificial thermal conductivity $\kappa_2^* \, \rho_{\infty}^{-1} u_{\infty}^{-1} c_v^{-1} d^{-1}$ (left) and artificial shear viscosity $\mu^* \rho_{\infty}^{-1} u_{\infty}^{-1} d^{-1}$ (right) fields for the hypersonic cylinder. Note $\kappa_2^*$ is non-zero in a small region downstream the cylinder. This corresponds to a sharp thermal feature that cannot be stabilized with a {\it shock capturing only} approach.}\label{f:avCylinder}
 \end{figure}

Figure \ref{f:avCylinder} shows a snapshot of the artificial thermal conductivity $\kappa_2^*$ and artificial shear viscosity $\mu^*$ fields. Both viscosities vanish in the shock wave. Unlike in the strong-vortex/shock-wave interaction problem, the artificial thermal conductivity is non-zero in a small region downstream the cylinder. This corresponds to a strong thermal gradient that cannot be stabilized with a {\it shock capturing only} approach. Indeed, removing $\kappa_2^*$ from the model leaded to the simulation breakdown. The use of artificial thermal conductivity stabilizes this feature without affecting the shock wave. This exemplifies the need to stabilize other under-resolved sharp features than shock waves for the simulation of high Reynolds, high Mach number unsteady flows.


\subsection{\label{s:T106C}Transonic T106C low-pressure turbine}

\subsubsection{Case description and numerical discretization}

We consider next the three-dimensional transonic flow over the T106C linear low-pressure turbine (LPT) in off-design conditions. The isentropic Reynolds and Mach numbers on the outflow are $Re_{2,s} = 100,817$ and $M_{2,s} = 0.987$, whereas the angle between the inflow velocity and the longitudinal direction is $\alpha_{1} = 50.54 \ \textnormal{deg}$. The extrusion length in the spanwise direction is 10\% of the blade chord $c_b$. The computational mesh consists of 712,080 isoparametric tetrahedral elements and the time-step size is $\Delta t = 6.94 \cdot 10^{-3} \ c_b \, \sqrt{\rho_{t,1} / p_{t,1}}$, where $p_{t,1}$ and $\rho_{t,1}$ are the inlet stagnation pressure and inlet stagnation density. Third-order HDG and DIRK(3,3) schemes are used for the discretization. A no-slip, adiabatic wall boundary condition is imposed on the blade surface, and a characteristics-based, non-reflecting boundary condition \cite{Fernandez:17a} is used on inflow and outflow. Periodicity is imposed on the tangential and spanwise directions.

\subsubsection{Numerical results}

\begin{figure}
\centering
\includegraphics[width=0.9\textwidth]{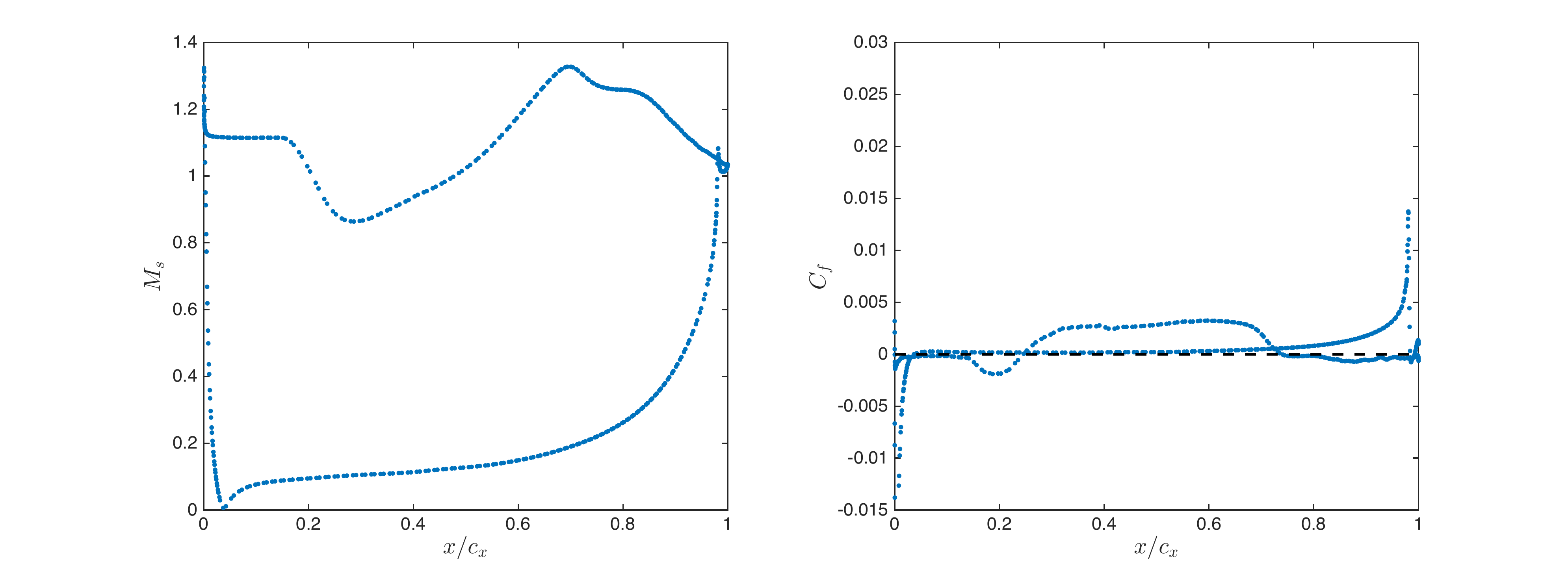}
\caption{\label{M2s_Cf} Isentropic Mach number $M_s$ (left) and skin friction coefficient $C_f$ (right) on the T106C low-pressure turbine blade. The stagnation pressure at inlet is used for non-dimensionalization of the skin friction coefficient.}
\end{figure}

\begin{figure}
\centering
\includegraphics[width=1.0\textwidth]{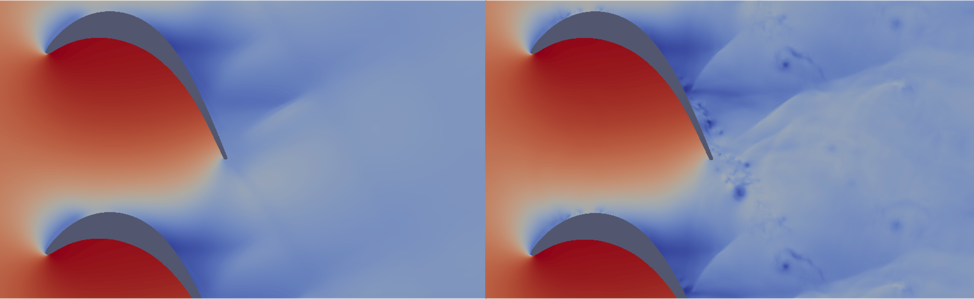}
\hfill \vspace{0.1mm} \includegraphics[width=1.0\textwidth]{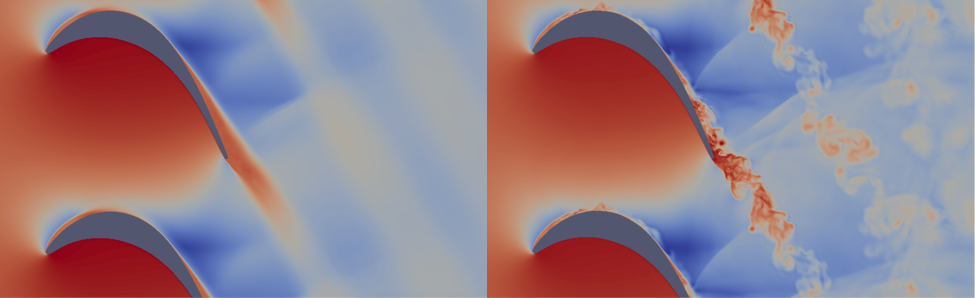}
\hfill \vspace{0.1mm} \includegraphics[width=1.0\textwidth]{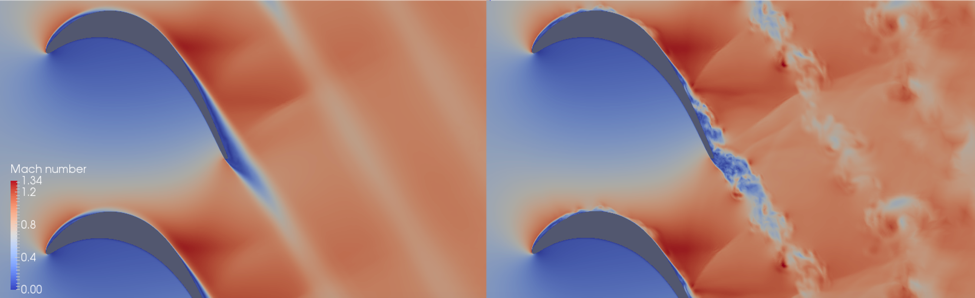}
\caption{\label{ptM} Pressure (top), temperature (center) and Mach number (bottom) fields for the transonic T106C LPT. Time-averaged and instantaneous fields are shown on the left and right images, respectively. The unsteady shocks involved are resolved within one element.}
\end{figure}

\begin{figure}
\centering
\includegraphics[width=1.0\textwidth]{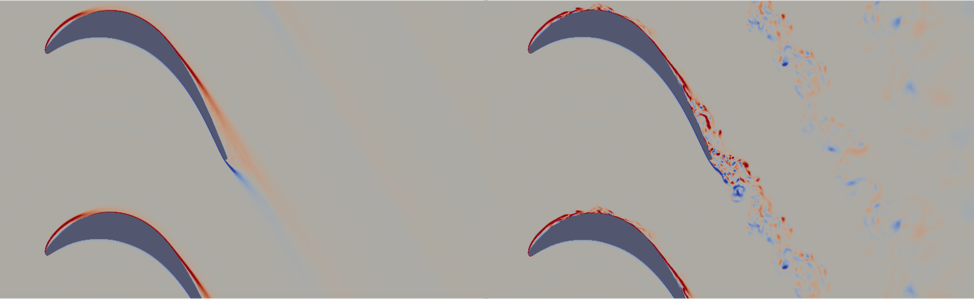}
\caption{\label{vorticity} Time-averaged (left) and instantaneous (right) spanwise vorticity fields for the transonic T106C LPT. The shock capturing method has a negligible impact on the vortical structures across the shock.}
\end{figure}

Figure \ref{M2s_Cf} shows the time- and spanwise-averaged isentropic Mach number (left) and skin friction coefficient (right) on the blade surface. The stagnation pressure at inlet $p_{t,1}$ is for non-dimensionalization of the skin friction coefficient. The time-averaged (left) and instantaneous (right) pressure, temperature and Mach number fields are shown in Figure \ref{ptM}. Several unsteady shocks that oscillate around a baseline position are present in this flow, as illustrated by the smoother shock profiles in the average fields compared to the instantaneous fields. These unsteady shocks are resolved within one element. Also, from the spanwise vorticity fields in Figure \ref{vorticity}, the shock capturing method has a negligible impact on the vortical structures across the shock. As discussed before, this is justified by the minor role of the bulk viscosity on the vorticity equation and will be further supported by the numerical results in Sections \ref{s:TGV} and \ref{s:CIT}. Finally, we emphasize that both $\kappa_2^*$ and $\mu^*$ are necessary to stabilize sharp thermal and shear features in this flow, such as the strong thermal gradient when the pressure and suction side boundary layers merge after the trailing edge.


%

\subsection{\label{s:TGV}Inviscid Taylor-Green vortex}


\subsubsection{Case description and numerical discretization}

The goal of this test problem is to examine the impact of the shock capturing method on the `solution quality' for severely under-resolved, shock-wave-free turbulence. The dissipation of kinetic energy, vortical structures and acoustic waves due to the shock capturing method is investigated. Not dissipating these features is critical to accurately simulate turbulent flows away from shocks. To this end, we perform large-eddy simulation of the inviscid Taylor-Green vortex (TGV) \cite{Taylor:37}. The TGV problem describes the evolution of an inviscid fluid in a cubic domain $\Omega = [-L \pi, L \pi)^3$ with triple periodic boundaries, starting from the smooth initial condition
\begin{equation}
\label{initialCondTGV}
\begin{split}
\rho & = \rho_0 , \\
v_1 & = V_0 \sin \Big( \frac{x}{L} \Big) \cos \Big( \frac{y}{L} \Big) \cos \Big( \frac{z}{L} \Big) , \\
v_2 & = - V_0 \cos \Big( \frac{x}{L} \Big) \sin \Big( \frac{y}{L} \Big) \cos \Big( \frac{z}{L} \Big) , \\
v_3 & = 0 , \\
p & = P_0 + \frac{\rho_0 \, V_0^2}{16} \ \bigg( \cos \Big( \frac{2x}{L} \Big) + \cos \Big( \frac{2y}{L} \Big) \bigg) \ \bigg( \cos \Big( \frac{2z}{L} \Big) + 2 \bigg) , 
\end{split}
\end{equation}
where $\rho_0$, $V_0$ and $P_0$ are positive constants, and $\bm{v} = (v_1, v_2, v_3)$ denotes the velocity vector. The reference Mach number is $M_0 = V_0 / c_0 = 0.1$, where $c_0$ is the speed of sound at temperature $T_0 = P_0 / (\gamma - 1) \, c_v \, \rho_0$. This completes the non-dimensional description of the problem. The flow is nearly incompressible and shock-wave free. The large-scale eddy in the initial condition leads to smaller and smaller structures through vortex stretching, until the vortical structures eventually break down and the flow transitions to turbulence\footnote{While no temporal chaos (chaotic attractor) exists in the inviscid Taylor-Green vortex, we use the term `turbulence' to refer to the phase of spatial chaos (spatial decoherence) that takes place after $t \approx 8 - 9 \, L / V_0$.}. Due to the lack of viscous dissipation, the smallest turbulent length and time scales in the exact solution become arbitrarily small as time evolves.


Third-order EDG and DIRK schemes are used for the discretization of the Euler equations. The computational domain is partitioned into a uniform grid with $64 \times 64 \times 64$ hexahedra; which leads to severe spatial under-resolution for this problem. The time-step size is $\Delta t = 3.68 \cdot 10^{-2} \, L / V_0$ and the numerical solution is computed from $t_0 = 0$ to $t_f = 10 \, L / V_0$. 
Three different phases can be distinguished in the simulation. Before $t \approx 4 \, L / V_0$, the flow is laminar and with no subgrid scales. This is followed by an under-resolved laminar phase that lasts until $t \approx 7-9 \, L / V_0$. From then on, the flow is turbulent and under-resolved.

\subsubsection{Numerical results}

\begin{figure}
\centering
\includegraphics[width=1.0\textwidth]{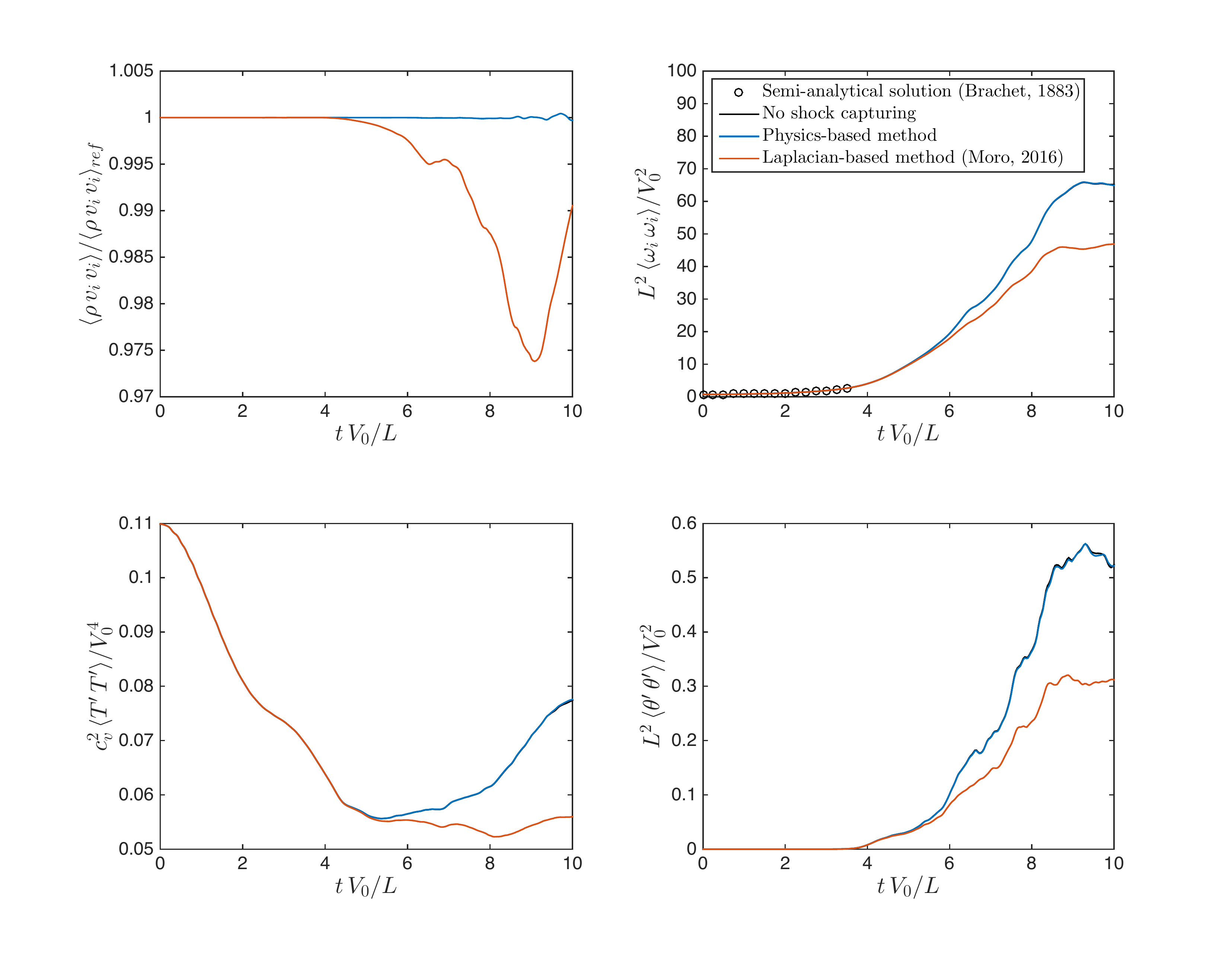}
\caption{\label{TGV_timeEvolutions} Temporal evolution of mean kinetic energy, mean-square vorticity, temperature variance and dilatation variance for the TGV problem. The ref subscript denotes the reference solution with no shock capturing, and $\langle \, \cdot \, \rangle$ denotes spatial averaging.}
\end{figure}


\begin{figure}
\centering-\includegraphics[width=1.0\textwidth]{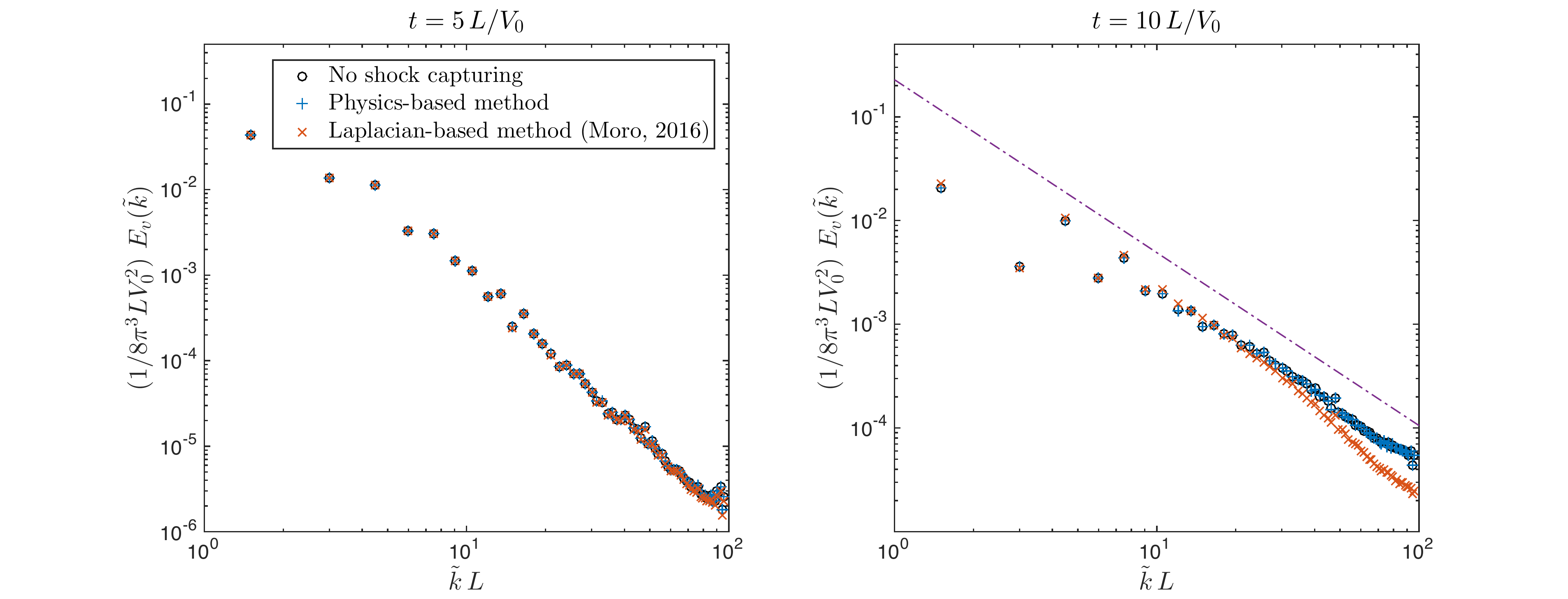}
\caption{\label{f:TGV_kinEnSpec} Kinetic energy spectrum at times $t = 5 \, L / V_0$ (left) and $t = 10 \, L / V_0$ (right) for the TGV problem. The Nyquist wavenumber of the grid is $\tilde{k}_N = 96 / L$. The theoretical $-5/3$ slope of decay of the inertial range of turbulence is shown in purple on the right figure.}
\end{figure}

The results with our physics-based method are compared to the results with no shock capturing (reference solution) and with the Laplacian-based method presented in \cite{Moro:2016}. 
The goal of the comparison with the Laplacian method in this and in the next test problem is to illustrate the importance of accurately detecting sharp features, as well as using only the physical viscosity that is required to stabilize the sharp feature, for large-eddy simulation of turbulent flows.

Figure \ref{TGV_timeEvolutions} shows the temporal evolution of the mean kinetic energy, mean-square vorticity, variance of temperature and variance of dilatation. The semi-analytical solution for the mean-square vorticity by Brachet {\it et al.} \cite{Brachet:1983} is shown as well. While $\langle \theta \rangle = 0$ in the exact solution due to periodicity in all directions, where $\langle \, \cdot \, \rangle$ denotes spatial averaging, we note this does not hold exactly, and thus variance of dilatation and mean-square dilatation are different $\langle \theta' \, \theta' \rangle \neq \langle \theta \, \theta \rangle$, in the discrete solution. 
When the flow is well-resolved and with no subgrid scales, i.e.\ before $t \approx 4 \, L / V_0$, both shock capturing methods agree with the reference solution; which in turn matches the semi-analytical data for the mean-square vorticity. As subgrid scales appear and the simulation becomes under-resolved, the physics-based method continues to have a minor impact on the numerical solution, whereas the Laplacian-based method dissipates all the compressible modes (i.e.\ vortical, entropy and acoustic modes). 
The superior performance of the physics-based method is mainly due to improved detection of sharp features by the sensors in Section \ref{s:sensors}. We emphasize that, while some small oscillations can be expected in under-resolved computations, particularly with high-order methods, the sensors should activate only for sharp features that cannot be resolved with the grid resolution and may lead to numerical instability.

Figure \ref{f:TGV_kinEnSpec} shows the one-dimensional kinetic energy spectrum at times $t = 5 \, L / V_0$ (left) and $t = 10 \, L / V_0$ (right). Note the Laplacian viscosity damps mostly the smallest resolved scales, as illustrated by the lower kinetic energy near the Nyquist wavenumber of the grid at $t = 10 \, L / V_0$. The Nyquist wavenumber is defined as $\tilde{k}_N = \pi / \hslash$, where $\hslash$ is the distance between high-order nodes, and corresponds to the smallest resolvable scales. The larger damping of the small scales is consistent with the second-order behavior of the Laplacian operator in wavenumber space, i.e.\ the decay rate of a signal is proportional to the square of its wavenumber.


\subsection{\label{s:CIT}Compressible isotropic turbulence}



\subsubsection{Case description and numerical discretization}

The goal of this test case is to investigate the robustness and the impact of the shock capturing method on the numerical solution for under-resolved compressible turbulence simulations. To this end, we consider the decay of compressible, homogeneous, isotropic turbulence with eddy shocklets \cite{Lee:1991}. The problem domain is a cube $\Omega = [-L \pi , L \pi)^3$ with triple periodic boundaries. The initial density, pressure and temperature fields are constant, and the initial velocity is solenoidal and with kinetic energy spectrum satisfying $E(\tilde{k}) \sim \tilde{k}^4 \, \exp [ - 2 \, (\tilde{k} / \tilde{k}_M)^2 ]$, where $\tilde{k}_M$ corresponds to the most energetic wavenumber and is set to $\tilde{k}_M = 4 / L$. The details of the procedure to generate the initial velocity field are described in \cite{Johnsen:2010}. 
The initial turbulent Mach number and Taylor-scale Reynolds number are
$$ \qquad M_{t,0} = \frac{\sqrt{ \langle v_{i,0} \, v_{i,0} \rangle}}{\langle c_0 \rangle} = 0.6 , \qquad \qquad Re_{\lambda,0} = \frac{\langle \rho_0 \rangle \, v_{rms,0} \, \lambda_0}{\langle \mu_0 \rangle} = 100 , $$
where the zero subscript denotes the initial value, $\langle \, \cdot \, \rangle$ denotes spatial averaging, and
$$ v_{rms} = \sqrt{\frac{\langle v_i \, v_i \rangle}{3}} , \qquad \qquad \lambda = \sqrt{\frac{ \langle v_1^2 \rangle }{ \langle (\partial_1 v_1)^2 \rangle}} $$
are the root mean square velocity and the Taylor microscale, respectively. Also, the shear viscosity is assumed to follow a power-law of the form
\begin{equation}
\mu = \mu_0 \, \bigg( \frac{T}{T_0} \bigg)^{3/4} . 
\end{equation}
This completes the non-dimensional description of the problem. Due to the imbalance in the initial condition, strong vortical, entropy and acoustic modes develop and persist throughout the simulation. Weak shock waves (eddy shocklets) appear spontaneously from the turbulent motions as well.
Third-order EDG and DIRK(3,3) schemes are used for the discretization. The computational domain is partitioned into a uniform $32 \times 32 \times 32$ Cartesian grid; which leads to severe spatial under-resolution for this problem. In order for the space discretization error to dominate the time discretization error, the time-step size is $\Delta t = 1.183 \cdot 10^{-2} \, \tau_0$, where $\tau_0 = \lambda_0 / v_{rms,0}$ denotes the initial eddy turn-over time. This corresponds to a CFL number based on the initial mean-square velocity of $v_{rms,0} \, \Delta t / h = 0.02$. The simulation is performed from $t_0 = 0$ to $t_f = 4 \, \tau_0$.

\subsubsection{Numerical results}

We present results for the physics-based method, the Laplacian-based method \cite{Moro:2016}, and a simulation with no shock capturing (reference solution). In addition, we consider the three following variations of the physics-based method. First, we set $Pr_{\beta}^* = 0.9$ instead of using Eq. \eqref{e:PrBetaStare2}. This will introduce some artificial thermal conductivity in shock waves through the term $\kappa_1^*$. Note that in the standard version of the model $Pr_{\beta}^* \gg 1$, and thus $\kappa_1^* \approx 0$, for the Mach numbers in this problem. Second, we take $k_{ \{ \kappa , \mu \} } = 0$ so that the terms $\kappa_2^*$ and $\mu^*$ vanish by construction. Third, we combine the two previous modifications and set $Pr_{\beta}^* = 0.9$ and $k_{ \{ \kappa , \mu \} } = 0$. 
We finally consider the direct numerical simulation (DNS) data from Hillewaert {\it et al.} \cite{Hillewaert:2016}. The grid resolution $\hslash$ in DNS is such that the 
P\'eclet number $Pe_{\hslash,0} = \langle \rho_0 \rangle \, v_{rms,0} \, \hslash / \langle \mu_0 \rangle$ is approximately $3.3$. 
While this suffices to stabilize the shock waves, it may not suffice to accurately resolve them and it is therefore unclear whether the DNS results are grid converged. 
Some differences between unfiltered DNS solutions computed with a finite-volume code and a DG code are indeed reported in \cite{Hillewaert:2016}.

Figure \ref{fig1_CIT} shows the temporal evolution of the mean-square velocity and vorticity, as well as the variance of temperature and dilatation, for all the methods considered. Since it is not obvious from these figures, we note that the two solutions with $Pr_{\beta}^* = 0.9$ display the same time evolutions, and the same is true for the two solutions with $Pr_{\beta}^*$ as given by Eq. \eqref{e:PrBetaStare2}. The reference simulation is unstable and breaks down at $t \approx 0.450 \, \tau_0$. A time refinement study confirmed the breakdown occurs independently of the time-step size, and it is therefore due to the lack of stability in the spatial discretization with no shock capturing. 
As discussed in Section \ref{s:introduction}, the role of the shock capturing method is to stabilize sharp features while having a small impact on the resolved turbulence and acoustic waves, and it is the role of the implicit or explicit SGS model to account for the effect of the subgrid scales. 
Hence, the solution with shock capturing should remain as close as possible to the reference solution without shock capturing, whenever the latter is stable. Note that no agreement with the DNS solution is expected {\it a priori} due to under-resolution, especially for quantities involving spatial derivatives of the numerical solution. 
The results in Figure \ref{fig1_CIT} can be summarized as follows:
\begin{itemize}
\item Except for dilatation variance, the physics-based method agrees with the reference solution before the latter breaks down, that is, it does not affect the numerical solution when no stabilization is required.
\item The dilatation in the reference solution suffers from severe Gibbs oscillations before the crash of the simulation, and this is in turn responsible for the breakdown. The physics-based method stabilizes the scheme by damping the Gibbs oscillations in dilatation. On the one hand, it does so without affecting the other compressible modes (i.e.\ the vortical and entropy modes). On the other hand, the damping of acoustic modes is excessive at that time compared to the DNS data. As discussed previously, it is unclear whether the DNS predictions, particularly of dilatation variance, are grid converged, and it is therefore challenging to infer additional conclusions from this figure.

\begin{figure}[t!]
\centering
\includegraphics[width=1.0\textwidth]{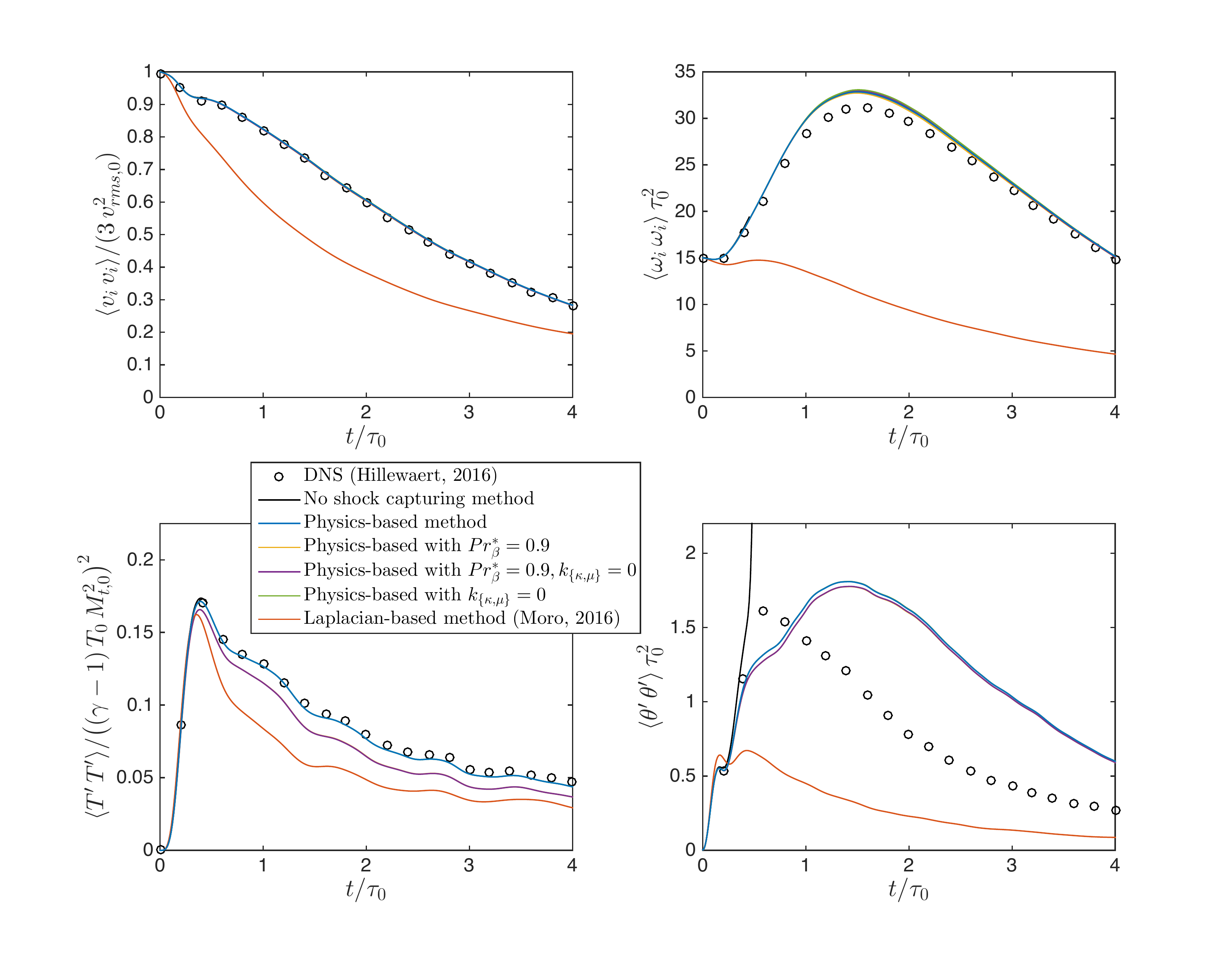}
\caption{\label{fig1_CIT} Temporal evolution of mean-square velocity, mean-square vorticity, temperature variance and dilatation variance for the compressible isotropic turbulence problem. The zero subscript denotes the initial value and $\langle \, \cdot \, \rangle$ denotes spatial averaging.}
\end{figure}

\item The physics-based method dissipates significantly less kinetic energy, vortical structures, temperature fluctuations and acoustic waves than the Laplacian-based method. The smaller impact on vortical, entropy and acoustic modes is critical for large-eddy simulation. 
In particular, shock stabilization through bulk viscosity, as opposed to shear viscosity or Laplacian viscosity, is key in order not to dissipate vortical structures across shock waves.
\item Overall, the physics-based method shows very good agreement with the DNS data, particularly when compared to the LES results with other shock stabilization methods \cite{Hillewaert:2016,Johnsen:2010} and despite the slightly lower resolution in the simulations in this paper.
\item If $Pr_{\beta}^*$ is set to $0.9$, the term $\kappa_1^*$ damps temperature fluctuations. If $Pr_{\beta}^*$ is as in Eq. \eqref{e:PrBetaStare2}, $\kappa_1^*$ vanishes and the scheme is still stable. This justifies the proposed form for $Pr_{\beta}^*$ in Eq. \eqref{e:PrBetaStare2}: Large for low and moderate Mach numbers (i.e.\ when $\kappa_1^*$ is not necessary) and asymptoting to $0.9$ for large Mach numbers (i.e.\ when it is required for stability and to obtain thermal and dynamic shock thicknesses of the same order).
\item No differences are observed between setting $k_{ \{ \kappa , \mu \} } = 1.0$ (the default value) and $k_{ \{ \kappa , \mu \} } = 0$. This shows that the thermal and shear sensors succeed to vanish in this problem, in which there are no sharp features other than shock waves.
\end{itemize}

\section{\label{s:conclusions}Conclusions}

We presented a physics-based shock capturing method for large-eddy simulation of turbulent flows. The proposed method performed robustly and provided sharp shock profiles for the transonic, supersonic, and hypersonic flows considered. Numerical results also indicated the method has a negligible impact on 
vortical structures, temperature fluctuations and dissipation of kinetic energy, both near and away from shocks. The impact on acoustic waves is negligible away from shocks, but some damping was observed near shocks. How to further improve the model to minimize the dissipation of acoustic modes across shocks is the subject of ongoing research. 

All the previous features are critical to enable robust and accurate large-eddy simulations of shock flows. From our experience, the key ingredients towards this end include: 
{\em i}) Shock stabilization through artificial bulk viscosity. This is an efficient mechanism to stabilize shock waves while having a negligible impact on the vortical structures across the shock. 
{\em ii}) Introducing also a small amount of artificial thermal conductivity in hypersonic shocks, required for stability and optimal shock resolution in hypersonic flows. 
{\em iii}) Accurate shock detection via dilatation- and vorticity-based sensors. 
{\em iv}) Artificial thermal conductivity and artificial shear viscosity to stabilize other under-resolved sharp features, such as strong thermal and shear layers. 
{\em v}) Accurate thermal and shear sensors that are active only in regions where the thermal and shear gradients are larger than possible with the grid resolution and may lead to numerical instability. 
{\em vi}) Smoothing the artificial viscosity fields to make them $\mathcal{C}^{0}$ continuous; which is critical for robustness. 
{\em vii}) Accounting for mesh anisotropy.

\section*{Acknowledgments}

The authors acknowledge the Air Force Office of Scientific Research (FA9550-16-1-0214), the National Aeronautics and Space Administration (NASA NNX16AP15A) and Pratt \& Whitney for supporting this effort. The first author also acknowledges the financial support from the Zakhartchenko and ``la Caixa'' Fellowships.

\appendix

\section{\label{s:DGnotation}Notation used for the hybridized DG discretization}

\subsection*{\label{s:FEmesh}Finite element mesh}

Let $\Omega \subset \mathbb{R}^d, \, 1 \leq d \leq 3$ be an open, connected and bounded physical domain with Lipschitz 
boundary $\partial \Omega$. We denote by $\mathcal{T}_h$ a collection of disjoint, non-singular, $p$-th degree curved elements $K$ that partition $\Omega$, and set $\partial \mathcal{T}_h := \{ \partial K : K \in \mathcal{T}_h \} $ to be the collection of the boundaries of the elements in $\mathcal{T}_h$. For an element $K$ of the collection $\mathcal{T}_h$, $F= \partial K \cap \partial \Omega$ is a boundary face if its $d-1$ Lebesgue measure is nonzero. For two elements $K^+$ and $K^-$ of $\mathcal{T}_h$, $F=\partial K^{+} \cap \partial K^{-}$ is the interior face between $K^+$ and $K^-$ if its $d-1$ Lebesgue measure is nonzero. We denote by $\mathcal{E}_h^I$ and $\mathcal{E}_h^B$ the set of interior and boundary faces, respectively, and we define $\mathcal{E}_h := \mathcal{E}_h^I \cup \mathcal{E}_h^B$ as the union of interior and boundary faces. Note that, by definition, $\partial \mathcal{T}_h$ and $\mathcal{E}_h$ are different. More precisely, an interior face is counted twice in $\partial \mathcal{T}_h$ but only once in $\mathcal{E}_h$, whereas a boundary face is counted once both in $\partial \mathcal{T}_h$ and $\mathcal{E}_h$.

\subsection*{Finite element spaces}

Let $\mathcal{P}_{k}(D)$ denote the space of polynomials of degree $k$ on a domain $D \subset \mathbb{R}^n$, let $L^2(D)$ be the space of Lebesgue square-integrable functions on $D$, and $\mathcal{C}^0(D)$ the space of continuous functions on $D$. Also, let $\bm{\psi}^p_K$ denote the $p$-th degree parametric mapping from the reference element $K_{ref}$ to an element $K \in \mathcal{T}_h$ in the physical domain, and $\bm{\phi}^p_F$ be the $p$-th degree parametric mapping from the reference face $F_{ref}$ to a face $F \in \mathcal{E}_h$ in the physical domain. We then introduce the following discontinuous finite element spaces
\begin{subequations}
\begin{alignat}{3}
& \bm{\mathcal{Q}}_{h}^k &&= \big\{\bm{r}_h \in [L^2(\mathcal{T}_h)]^{m \times d} \ : \ (\bm{r}_h \circ \bm{\psi}_K^p )  |_K \in [\mathcal{P}_k(K_{ref})]^{m \times d} \ \ \forall K \in \mathcal{T}_h \big\} , \\
& \bm{\mathcal{V}}_{h}^k &&= \big\{\bm{w}_h \in [L^2(\mathcal{T}_h)]^m \ : \ (\bm{w}_h \circ \bm{\psi}_K^p )|_K \in [\mathcal{P}_k(K_{ref})]^m \ \ \forall K \in \mathcal{T}_h \big\} , \\
& \bm{\mathcal{M}}_{h}^k  &&= \big\{ \bm{\mu}_h \in [L^2(\mathcal{E}_h)]^m \ : \ (\bm{\mu}_h \ \circ \ \bm{\phi}^p_F)|_F \in [\mathcal{P}^k(F_{ref})]^m \, \ \forall F \in \mathcal{E}_h , \ \textnormal{and} \ \bm{\mu}_h|_{\mathcal{E}^{\rm E}_h} \in [C^0(\mathcal{E}^{\rm E}_h)]^m \big\} , 
\end{alignat}
\end{subequations}
where $\mathcal{E}^{\rm E}_h$ is a subset of $\mathcal{E}_h$, and $m$ denotes the number of equations of the conservation law, i.e.\ $m=d+2$ for the Euler and Navier-Stokes systems. 
Note that $\bm{\mathcal{M}}_{h}^k$ consists of functions which are continuous on $\mathcal{E}^{\rm E}_h$ and discontinuous on $\mathcal{E}^{\rm H}_h := \mathcal{E}_h \backslash \mathcal{E}^{\rm E}_h$. Different choices of $\mathcal{E}^{\rm E}_h$ lead to different discretization methods that have different properties in terms of accuracy, stability, and number of globally coupled unknowns \cite{Fernandez:17a}. In particular, the Hybridizable DG (HDG), Embedded DG (EDG) and Interior Embedded DG (IEDG) methods are obtained by setting $\mathcal{E}^{\rm E}_h = \emptyset$, $\mathcal{E}^{\rm E}_h = \mathcal{E}_h$ and $\mathcal{E}^{\rm E}_h = \mathcal{E}_h^I$, respectively. Further discussion on this family of schemes is presented in \cite{Fernandez:17a,Fernandez:PhD:2018}.

It remains to define inner products associated with these finite element spaces. For functions $\bm{a}$ and $\bm{b}$ in $[L^2(D)]^m$, we denote $(\bm{a},\bm{b})_D = \int_{D} \bm{a} \cdot \bm{b}$  if $D$ is a domain in $\mathbb{R}^d$ and $\left\langle \bm{a},\bm{b}\right\rangle_D = \int_{D} \bm{a} \cdot \bm{b}$ if $D$ is a domain in $\mathbb{R}^{d-1}$. Likewise, for functions $\bm{A}$ and $\bm{B}$ in $[L^2(D)]^{m \times d}$, we denote $(\bm{A},\bm{B})_D = \int_{D} \mathrm{tr}(\bm{A}^T \bm{B})$  if $D$ is a domain in $\mathbb{R}^d$ and $\left\langle \bm{A},\bm{B}\right\rangle_D = \int_{D} \mathrm{tr}(\bm{A}^T \bm{B})$ if $D$ is a domain in $\mathbb{R}^{d-1}$, where $\mathrm{tr} \, ( \cdot) $ is the trace operator of a square matrix. We finally introduce the following inner products
\begin{equation*}
(\bm{a},\bm{b})_{\mathcal{T}_h} = \sum_{K \in \mathcal{T}_h} (\bm{a},\bm{b})_K, \qquad (\bm{A},\bm{B})_{\mathcal{T}_h} = \sum_{K \in \mathcal{T}_h} (\bm{A},\bm{B})_K, \qquad \left\langle \bm{a},\bm{b}\right\rangle_{\partial \mathcal{T}_h} = \sum_{K \in \mathcal{T}_h} \left\langle \bm{a},\bm{b}\right\rangle_{\partial K} . 
\end{equation*}

\section{\label{s:optimalKkappa}Theoretical estimate of the optimal value of $k_{\beta}$}

We present an estimate of the value of $k_{\beta}$ to optimally resolve a stationary normal shock wave with the grid resolution. First, let us define a modified viscosity $\tilde{\mu} = 4 \, \mu / 3 + \beta$ and a modified Prandtl number $\widetilde{Pr} = c_p \, \tilde{\mu} / \kappa$, where $(\beta , \kappa , \mu)$ are the sum of the physical and artificial viscosities. For a Newtonian, calorically perfect gas in thermodynamic equilibrium\footnote{Although these assumptions may not hold inside an actual shock wave, these are the physical models used for the numerical discretization and therefore those to be used to estimate the value of $k_{\beta}$.}, the entropy production across a stationary normal shock can be shown to be approximately given by
\begin{equation}
\label{deltaS1}
s_2 - s_1 \approx \frac{2}{\rho v \delta_s} \frac{\tilde{\mu} \frac{v_1^2}{T_1} \big( \frac{v_2}{v_1} - 1 \big)^2 + 2 \kappa \, \big( \frac{T_2}{T_1} - 1 \big)^2 \big( \frac{T_2}{T_1} + 1 \big) ^{-1}}{\big(\frac{T_2}{T_1}+1 \big)} , 
\end{equation}
where the subscripts $1$ and $2$ denote the upstream and downstream conditions, and $\delta_s$ is the dynamic thickness of the shock. We have assumed the dynamic and thermal thicknesses are of the same order $\delta_s \approx \theta_s$; which is the case for example if $\widetilde{Pr}$ is {{of}} order $1$. In addition, the following approximations have been used inside the shock
\begin{equation}
\label{e:approximations}
\frac{dv}{dx}(\bm{x}) \approx \frac{v_2 - v_1}{\delta_s} , \qquad \qquad \frac{d T}{dx}(\bm{x}) \approx \frac{T_2 - T_1}{\delta_s} , \qquad \qquad T(\bm{x}) \approx \frac{T_1 + T_2}{2} . 
\end{equation}
Alternatively, the entropy jump can be expressed in terms of the density and temperature ratios using Gibbs' equation, namely,
\begin{equation}
\label{deltaS2}
s_2 - s_1 = c_v \, \ln \bigg[ \frac{T_2}{T_1} \bigg( \frac{\rho_1}{\rho_2} \bigg) ^{\gamma-1} \bigg] . 
\end{equation}
Combining Equations \eqref{deltaS1} and \eqref{deltaS2}, it follows that
\begin{equation}
\label{eq1app}
\frac{\rho v \delta_s}{\bar{\mu}} \approx \frac{ 2 \gamma \, (\gamma - 1) \, M_{1}^2 \Big( \frac{\rho_1}{\rho_2} -1 \Big)^2 + 4 \gamma \, \widetilde{Pr}^{-1}    \Big(\frac{T_2}{T_1}-1\Big)^2 \Big(\frac{T_2}{T_1}+1\Big)^{-1} } { \Big( \frac{T_2}{T_1} + 1 \Big) \ln \Big[ \frac{T_2}{T_1} \Big(\frac{\rho_1}{\rho_2}
 \Big)^{\gamma-1} \Big] } =: \mathcal{F}(M_{1},\gamma,\widetilde{Pr}) . 
\end{equation}
Moreover, $\beta^* \gg \beta_f, \mu_f, \mu^*$ in a shock wave with our method, and thus $\tilde{\mu} \approx \beta^*$. Further assuming $Pr_{\beta}^*$ is set to be of order $1$, it follows that $\kappa^* \gg \kappa_f$, $\widetilde{Pr} \approx Pr_{\beta}^*$ and
\begin{equation}
\label{eq4app}
\beta^* \approx \frac{\rho v \delta_s }{\mathcal{F}(M_{1},\gamma,Pr_{\beta}^*)} . 
\end{equation}
Note that $Pr_{\beta}^* \approx 1$ in turn ensures the previous assumption $\delta_s \approx \theta_s$ holds.

 \begin{figure}[b!]
 \centering
 {\includegraphics[width=0.6\textwidth]{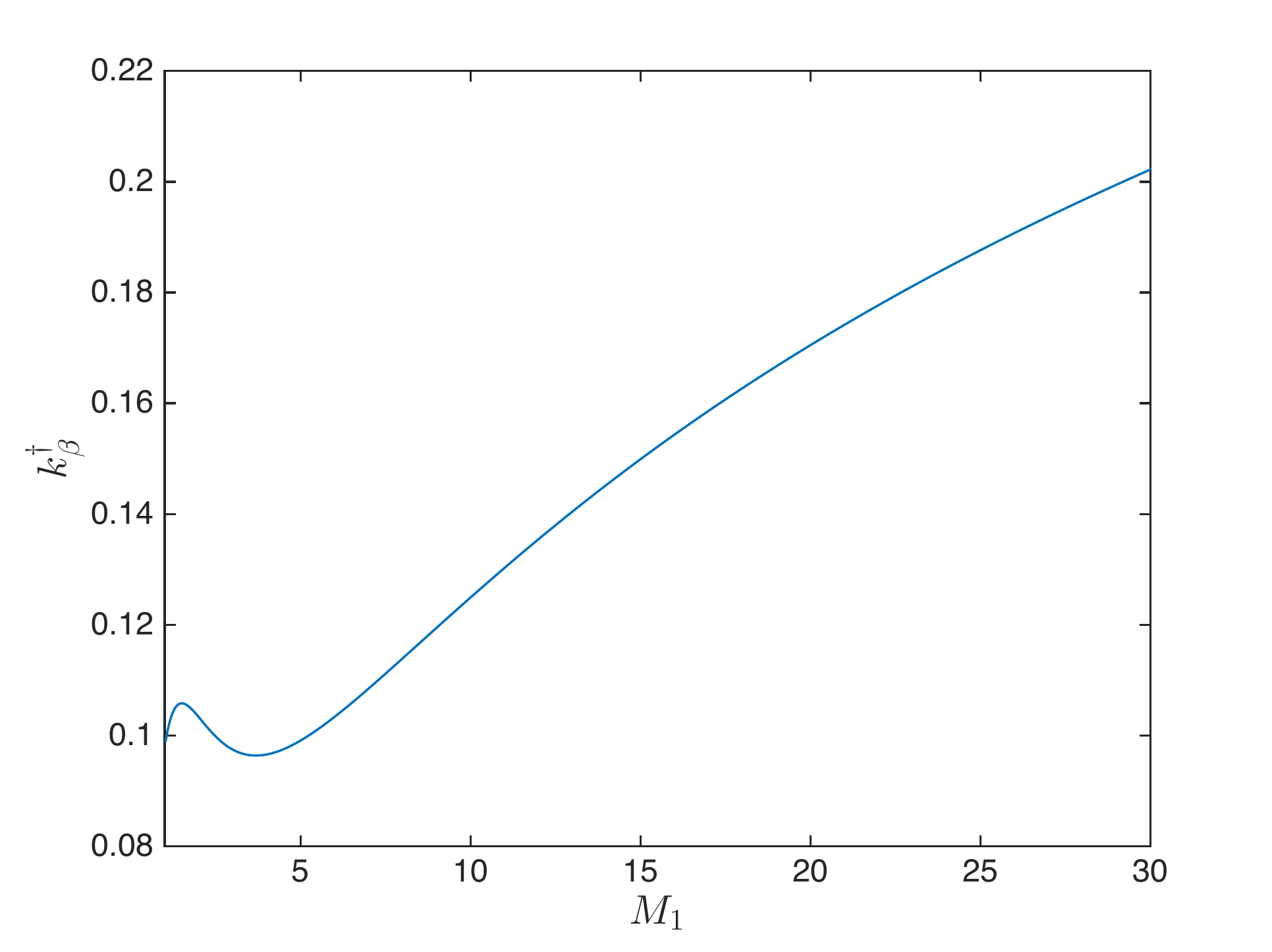}}
  \caption{Estimate of the value of $k_{\beta}$, denoted by $k_{\beta}^{\dagger}$, to optimally resolve a stationary normal shock wave. The case $\gamma = 1.4$ and $Pr_{\beta}^* = 0.9$ is shown.}\label{f:kBetaOptVsM1}
 \end{figure}

The artificial bulk viscosity in a stationary normal shock is
\begin{equation}
\label{eq2app}
\beta^* (\bm{x}) \approx k_{\beta} \frac{h_{\beta}^2}{k^2} \  \frac{\rho v}{\delta_s} \ \mathcal{H} \big( M (\bm{x}), M_{1} , \gamma \big) , 
\end{equation}
where
\begin{equation}
\label{eq3app}
\mathcal{H}(M, M_{1} , \gamma) := \frac{2 M_{1}^2 - 2}{(\gamma+1) \, M_{1}} \ \bigg( \frac{2}{2 + (\gamma-1) \, M_{1}^2} \, \Big( \gamma + \frac{1}{M^2} \Big) \bigg)^{1/2} . 
\end{equation}
Note that the artificial bulk viscosity is a function of the position $\bm{x}$ due to the $M(\bm{x})$ term. From Equations \eqref{eq4app} and \eqref{eq2app}, an estimate of the value of $k_{\beta}$ to optimally resolve the shock with the grid resolution, i.e.\ $\delta_s , \theta_s \approx h_{\beta} / k$, is given by
\begin{equation}
\label{k_betaSuggested}
k_{\beta}^{\dagger} = \Big[ \mathcal{F} \big( M_{1},\gamma,Pr_{\beta}^* \big) \ \mathcal{H} \big( M_1 , M_{1} , \gamma \big) \Big]^{-1} , 
\end{equation}
which needs to be used in conjunction with $Pr_{\beta}^* \approx 1$. Figure \ref{f:kBetaOptVsM1} plots $k_{\beta}^{\dagger}$ for the particular case of $\gamma = 1.4$ and $Pr_{\beta}^* = 0.9$. While $k_{\beta}^{\dagger}$ is between $0.1$ and $0.2$ for incident Mach numbers below $30$, our experience from numerical experiments is that $k_{\beta} \approx 1$ is required for stability in practice. The difference between the theoretical estimate and the value required in practice is mostly attributed to the approximations used in the derivation of $k_{\beta}^{\dagger}$. 
Numerical experiments also indicate that $Pr_{\beta}^* \approx 1$ is only necessary in practice for hypersonic shocks. This justifies the proposed form for $Pr_{\beta}^*$ in Eq. \eqref{e:PrBetaStare2}; which is large for low and moderate Mach numbers (i.e.\ when $\kappa_1^*$ is not necessary) and asymptotes to $0.9$ for large Mach numbers (i.e.\ when it is required to stabilize the scheme and to obtain thermal and dynamic shock thicknesses of the same order).


%
%
%

\end{document}